\documentclass[aip,rsi,reprint,amsmath,amssymb,floatfix]{revtex4-2}

\usepackage{graphicx}
\usepackage{dcolumn}
\usepackage{bm}
\usepackage{hyperref}
\usepackage[utf8]{inputenc}
\usepackage{textcomp}
\usepackage{sidecap}
\usepackage{placeins}
\usepackage[utf8]{inputenc}
\usepackage[T1]{fontenc}

\makeatletter
\def\@email#1#2{%
 \endgroup
 \patchcmd{\titleblock@produce}
  {\frontmatter@RRAPformat}
  {\frontmatter@RRAPformat{\produce@RRAP{*#1\href{mailto:#2}{#2}}}\frontmatter@RRAPformat}
  {}{}
}%
\makeatother
\begin{document}

\title{A precision apparatus for high harmonic spectroscopy in bulk solids}

\author{S. Mandal}
\author{P. Kumar}
\author{Z. Pi}
\author{H. Y. Kim}
\author{M. Zhan}
\author{E. Goulielmakis*}
\email{e.goulielmakis@uni-rostock.de}
\affiliation{Institut für Physik, Universität Rostock, Albert-Einstein-Straße 23--24, 18059 Rostock, Germany}

\date{\today}

\begin{abstract}
High harmonic generation (HHG) in solids has emerged as a powerful spectroscopic method for resolving ultrafast electron dynamics and band structure properties across a wide range of materials. However, quantitative HHG studies require instrumentation capable of delivering stable driving fields, precise crystal alignment, and broadband detection spanning the UV to the extreme ultraviolet (EUV). Here we present an integrated apparatus engineered specifically for high-accuracy, field-strength and orientation-dependent HHG measurements in bulk solids. The system incorporates dispersion-neutral intensity-control for few-cycle pulses, a vacuum HHG module with sub-micrometer and sub-degree sample positioning, and an imaging assembly that stabilizes the focal spot position and enables spatial filtering of the emitted harmonics. A synchronized dual-spectrometer scheme provides simultaneous UV/VUV and EUV radiation detection, while absolute electric field calibration is achieved through gas-phase attosecond streaking. Together, these capabilities establish a versatile and quantitatively reliable platform for solid-state HHG spectroscopy. The methodology is broadly adaptable to various laser sources and material classes, and supports future efforts aimed at reconstructing valence-electron potentials, tracking strong-field dynamics, and mapping electronic structure with sub-cycle temporal resolution.
\end{abstract}

\maketitle

\section{Introduction}

High harmonic generation (HHG) in solids has evolved into a powerful
spectroscopic method for probing the electronic structure across a broad
range of materials. Since the first observation of HHG in bulk
semiconductors under intense mid-infrared fields\cite{Ghimire2011}, the
technique has been extended to dielectrics\cite{Luu2015,Garg2016},
amorphous media\cite{Jurgens2020}, two-dimensional
semiconductors\cite{Liu2017}, gapless materials such as
graphene\cite{Yoshikawa2017}, and strongly correlated and topological
systems where HHG provides access to many-body dynamics, phase
transitions, and surface-state responses\cite{Silva2018,Murakami2018,Bauer2018,Baykusheva2021}. Across
these systems, the emission characteristics -- cutoff energies, harmonic
yields, and angular anisotropies -- encode information on band structure,
multiband coupling, Berry curvature, and valence-electron
potentials\cite{You2017,Lanin2017,Luu2018,Lakhotia2020,You2018}. These advances establish HHG as an
emerging metrological tool for quantitative solid-state spectroscopy
extending from the infrared to the EUV.

Such quantitative studies, however, impose stringent requirements for
adequate instrumentation. HHG yields from thin crystalline samples are
highly sensitive to sample inhomogeneities and beam-pointing drifts,
making field-strength scans and angular scans particularly vulnerable to
systematic errors. Accurate extraction of electronic-structure
information calls for (i) stable, dispersion-free control of the driving
electric-field amplitude especially for few-cycle pulses, (ii) precise
and reproducible crystal rotation about a well-defined axis, and (iii)
simultaneous detection of harmonics across extended spectral ranges
spanning from the visible to the EUV. Moreover, accurate determination
of the driving-field amplitude -- necessary for quantitative
reconstruction of material properties -- requires dedicated calibration
methods\cite{Goulielmakis2004}.

In this work, we present an integrated apparatus engineered to meet
these requirements for precision HHG spectroscopy in solids. The
apparatus is particularly optimized for few-cycle driving fields in the
near-infrared and visible parts of the spectrum. The system combines a
dispersion-neutral intensity-control module, a vacuum HHG platform
equipped with a custom multi-axis goniometer for sub-micrometer
positioning and sub-degree rotational accuracy, an imaging and spatial
filtering assembly that ensures stable focal spot interrogation during
field and angle scans, and a dual-spectrometer system enabling
simultaneous UV/VUV and EUV radiation detection. Absolute calibration of
the driving electric field is performed via gas-phase attosecond
streaking\cite{Goulielmakis2004}. Together, these capabilities constitute a
versatile platform for accurate field-strength- and
orientation-dependent HHG measurements, supporting a wide range of
measurements in high harmonic spectroscopy of solids.

\begin{figure*}
\centering
\includegraphics[width=2.1\columnwidth,keepaspectratio,height=1\textheight]{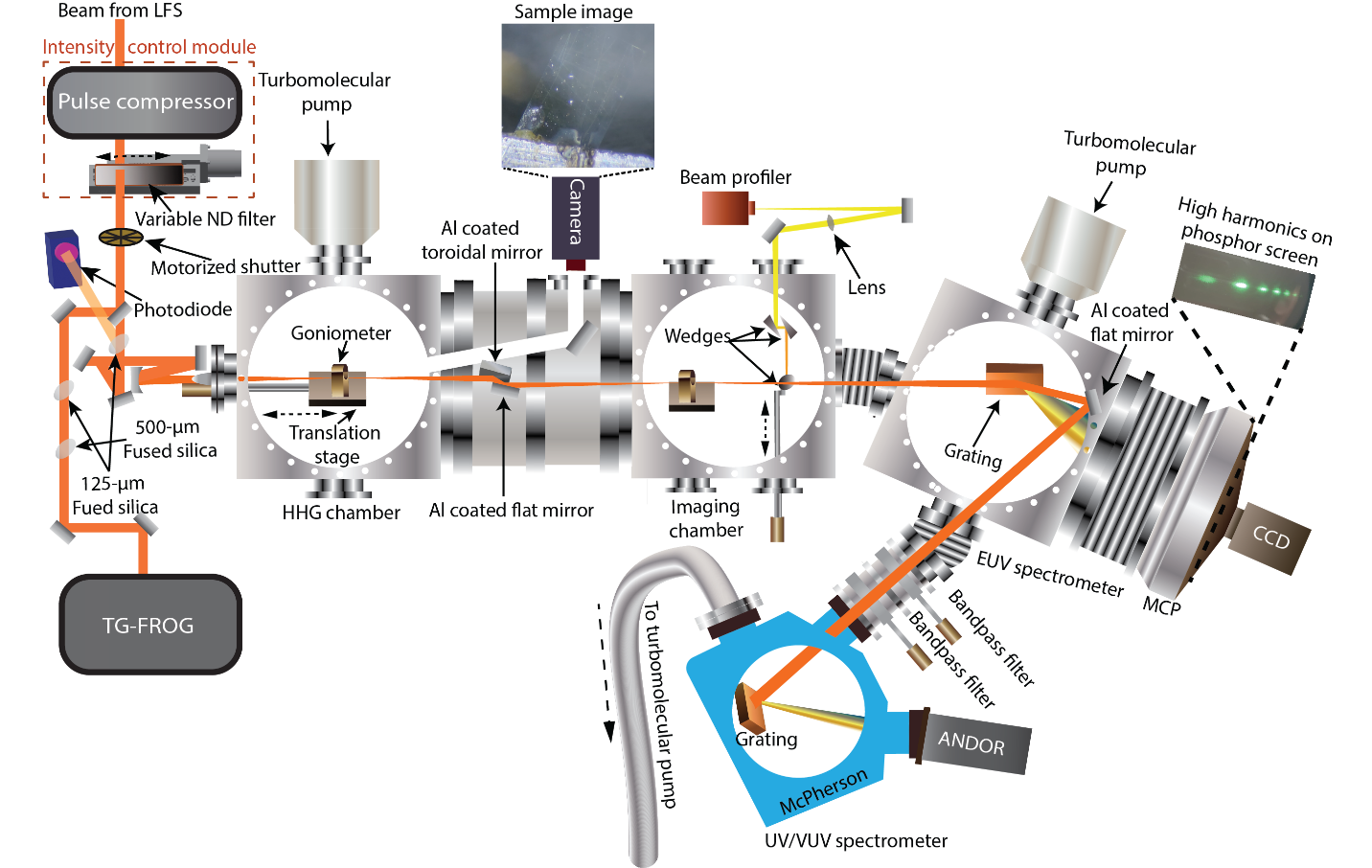}
\caption{Schematic of the high harmonic spectroscopy apparatus.}
\label{fig:1}
\end{figure*}

\section{OVERVIEW OF THE EXPERIMENTAL APPARATUS}

The experimental apparatus (Fig.~\ref{fig:1}) consists of four main modules: (a)
the intensity control module -- for precise tuning of the driving pulse
intensity; (b) the high harmonic generation module -- which facilitates
the positioning and rotation of the crystalline specimen to be
investigated at the laser beam focus; (c) the imaging and spatial
filtering assembly -- which enables imaging of the high harmonic source
onto the entrance plane of the spectrometers, enabling also a spatial
selection of harmonics emanating from specific regions of the
high harmonic beam; and (d) the detection system, where two
spectrometers allow simultaneous detection of the generated harmonics in
the UV/VUV and EUV ranges. Except module (a), which operates in ambient
air, modules (b), (c) and (d) are housed within three interconnected
chambers (Fig.~\ref{fig:1}) maintained at a vacuum level of 10\textsuperscript{-6}
mbar, allowing an absorption-free propagation of VUV and EUV harmonics
from the source to the detection system. The apparatus is driven by a
light field synthesizer (LFS)\cite{Wirth2011,Hassan2012,Hassan2016} which, for the
applications described here, provides few-cycle pulses in the visible
spectral range centered at 2.0 eV and the near infrared (NIR) region
centered at 1.5 eV.

In the following sections we focus on describing the technical details
and purpose of each of the modules mentioned above and explain how each
of these accomplishes key goals for attaining precision measurements of
high harmonic emission from bulk solids.

\subsection{Intensity control module}

Studies of high harmonic generation from bulk solids have shown that
valuable information about the electronic structure of a crystal is
encoded both in the harmonic yield\cite{Luu2015,Lakhotia2020} as well as the
cutoff energy as a function of the applied electric field amplitude.
Therefore, a high harmonic spectroscopy apparatus geared towards the
precise characterization of the emitted harmonics must incorporate a
dedicated pulse intensity control module. For ultrashort pulses, this
module must vary the intensity without altering their temporal or
spatial characteristics, in order to minimize measurement
uncertainties.

In principle, accurate, dispersion-free intensity control of few-cycle
pulses could be implemented by spatially varying neutral density (ND)
reflectors (filters), whose reflectivity gradient is negligible over the
laser beam size. Yet the spatial translation of the reflector, necessary
to adjust the pulse intensity to a specific setting, unavoidably affects
the beam alignment. While in gas phase high harmonic generation, a
moderate misalignment of the beam path and a corresponding lateral
translation of the focal spot in the gas jet is of minor significance,
it can severely affect measurement precision in bulk solids. This is
because bulk solid samples (especially very thin ones) suffer surface
imperfections such as digs, scratches, as well as non-uniform degree of
crystallinity. Consequently, yield of laser-driven high harmonics from
thin solid specimen is rather sensitive to the selected spot on the
sample, on which the laser beam is focused to.

\begin{figure}[!tb]
\centering
\includegraphics[width=0.95\columnwidth,keepaspectratio,height=1\textheight]{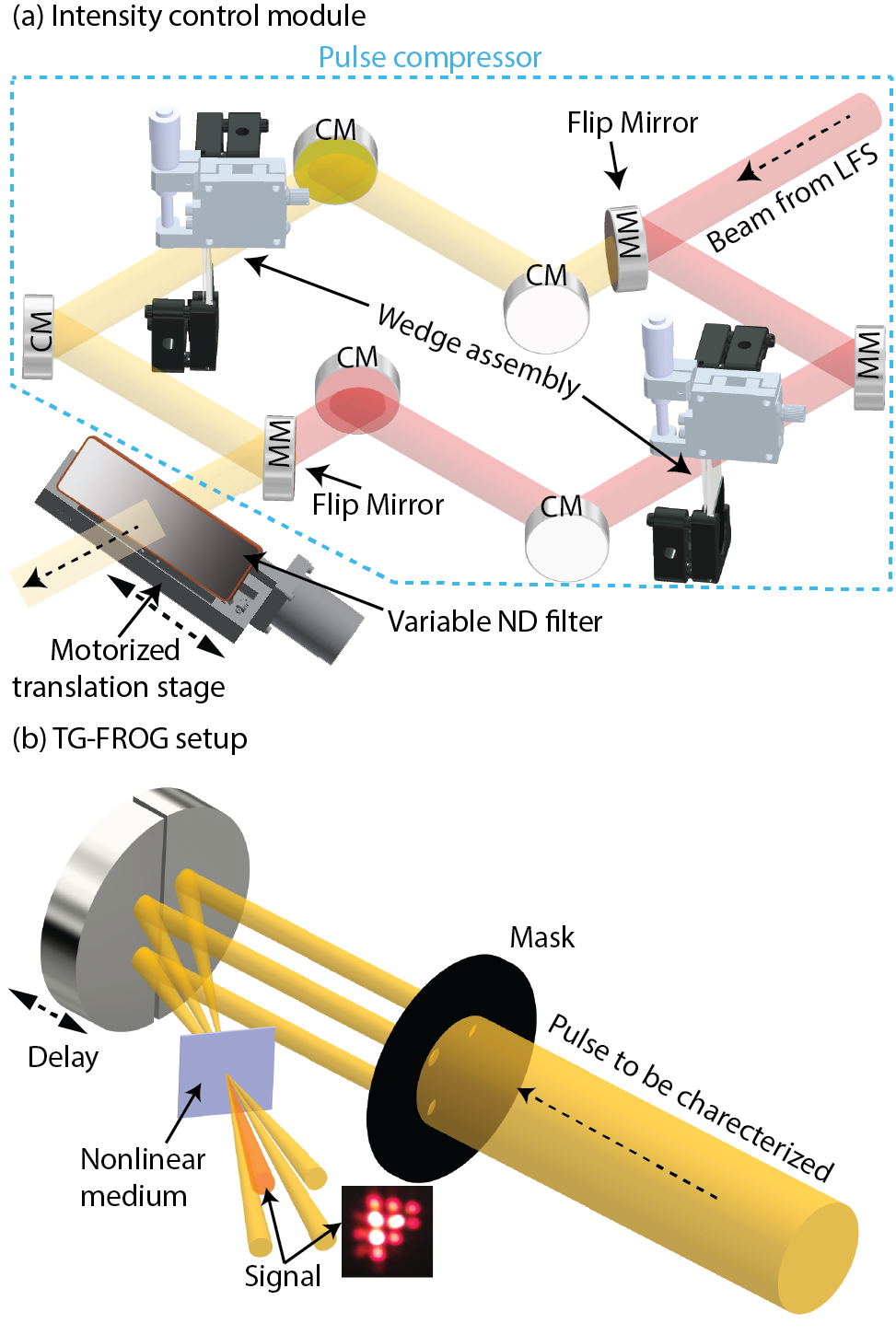}
\caption{Intensity control module and TG-FROG. (a) Schematic of the visible (yellow) and NIR (red) pulse compressor with the variable ND filter. Pulses from LFS are negatively chirped by chirped mirrors (CM) and dispersed by a pair of wedges in each individual path. Metal-coated mirrors are marked as MM. (b) Schematic of the TG-FROG setup for temporal characterization of the pulse (see also reference \onlinecite{Sweetser1997}).}
\label{fig:2}
\end{figure}

The intensity control module implemented in this apparatus minimizes
angular deviations during the adjustment of the pulse intensity and
ensures that during a pulse-intensity scan, harmonics are emerging from
the same spot of the studied specimen. Technically, the intensity control
module (Fig.~\ref{fig:2}(a)) is implemented using (a) a set of chirp mirrors to
negatively chirp the few-cycle pulses, and (b) a continuously variable,
linear ND filter in transmission to adjust their energy. In this
implementation a rectangular (25 mm $\times$ 100 mm) continuously variable ND
filter having a thickness of $\sim$2.2 mm is used. Owing to the
chirp-induced pulse intensity suppression, nonlinear effects in the bulk
of ND filter are minimized. Moreover, to enable a uniform attenuation of
the pulse intensity across the beam profile, which can be compromised by
the spatial gradient of the ND filter, the diameter of the laser beam
was kept lower than $\sim$3.5 mm. The intensity control
module is completed by (c) a set of Brewster-angled, fused silica wedges
that allow fine-tuning of the pulse dispersion. Using the module, the
pulses acquire their minimum duration when they reach the high harmonic
target specimen, as described later in this work. As we are dealing with
pulses at two different bands the module is comprised by two paths. \textit{Path
1} is designed for pulses in the spectral range 500 nm -- 700 nm, while
\textit{path 2} for pulses in the range 700 nm -- 1100 nm. In \textit{path 1} dispersion control is implemented using three chirped mirrors (-70
fs\textsuperscript{2} per reflection, 500 nm -- 700 nm bandwidth), while in \textit{path 2} a pair of chirped mirrors (-70 fs\textsuperscript{2} per reflection, 700 nm -- 1100 nm bandwidth) were necessary.
Routing of the beam through \textit{path 1} or \textit{path 2} is possible via flip mirrors (Fig.~\ref{fig:2}(a)) placed at the entrance and exit of the module. The wedge-pair, mounted on a micrometer translation stage, allows
fine-tuning of the dispersion of the pulses before reaching the high
harmonic module. The pulse intensity control not only warrants a smooth
and misalignment-free variation of the intensity but also ensures
minimal effects on the beam properties in the focus in comparison to the
aperture-based \cite{Nefedova2018,Chevreuil2021} intensity control extensively
used in the past for few-cycle pulse intensity control.

For the temporal characterization of the few-cycle pulses in our
apparatus a compact TG-FROG\cite{Sweetser1997} setup is placed at the
exit of the pulse intensity control module. To ensure that the pulses
characterized in the FROG setup accurately represent those used for
high harmonic generation, we matched the optical paths that guide the
beam to the HHG chamber and to the TG-FROG (Fig.~\ref{fig:1}). The schematic of
the FROG apparatus is shown in Fig.~\ref{fig:2}(b).

\sidecaptionvpos{figure}{c}
\begin{SCfigure*}
\includegraphics[width=1.5\columnwidth,keepaspectratio,height=0.8\textheight]{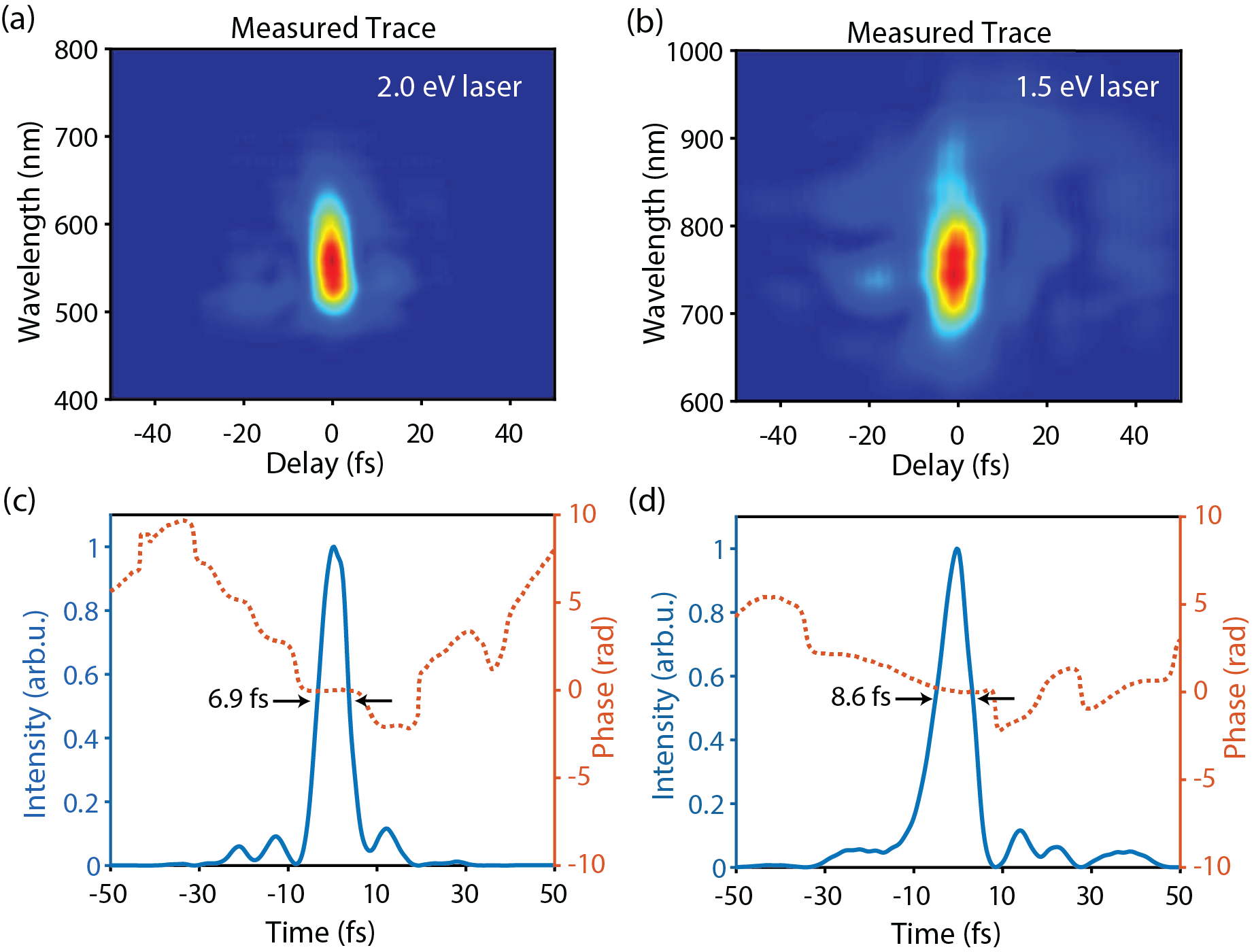}
\caption{\protect Temporal characterization of the pulses used in high harmonic generation experiments. (a), (b) Measured TG-FROG spectrograms of the driving pulses centered at 2 eV and 1.5 eV respectively. (c), (d) Retrieved temporal intensity profile (blue) and phase (red). The evaluated pulse durations at the FWHM are $\tau_{\mathrm{VIS}}$ $\simeq$ 6.9 fs and $\tau_{\mathrm{NIR}}$ $\simeq$ 8.6 fs respectively.}
\label{fig:3}
\end{SCfigure*}

The temporal intensity profiles and phases of the pulses are retrieved
by reconstruction of the recorded FROG spectrograms using a commercial
software (Femtosoft Technologies). Figures \ref{fig:3}(a) and \ref{fig:3}(b) show the
measured FROG spectrograms of visible and NIR pulses respectively. The
retrieved temporal intensity and corresponding temporal phase of the
pulses are shown in Fig.~\ref{fig:3}(c) and \ref{fig:3}(d). The evaluated durations (FWHM)
of the pulses are $\tau$ $\simeq$ 6.9 fs and $\tau$ $\simeq$ 8.6 fs respectively.

\subsection{High harmonic generation module}

\begin{figure*}
\centering
\includegraphics[width=1.6\columnwidth,keepaspectratio,height=0.8\textheight]{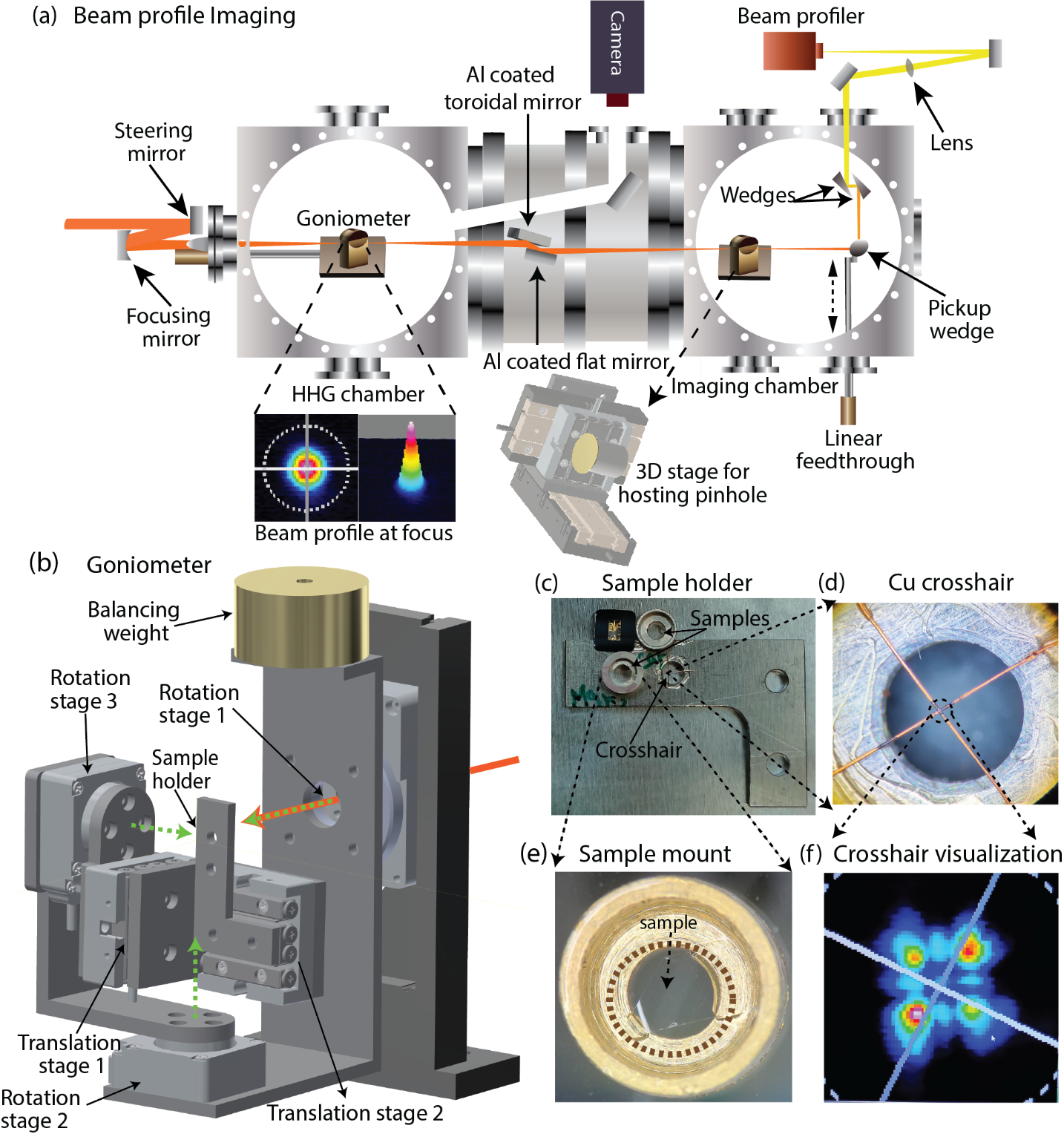}
\caption{(a) Overview of the f-to-2f imaging system used for aligning the rotation axis of the crystal sample along the laser propagation axis (orange solid arrow). (b) Multi-axis goniometer with the sample holder. The green dashed arrows denote the axes of rotation of each rotation stage (c) Photograph of the sample holder with three mounted samples and a crosshair. (d), (e) Photograph of the crosshair and sample respectively. (f) The crosshair imaged at the focal plane.}
\label{fig:4}
\end{figure*}

The driving pulses exiting the intensity control module are routed to
the entrance of the high harmonic generation module through a series of
reflections from silver-coated steering mirrors. A concave silver-coated
mirror (f = 40 cm) is used to focus the beam at approximately the center
of the chamber (Fig.~\ref{fig:1}) maintained in vacuum
($\sim1$0\textsuperscript{-6} mbar). The pulses enter the
chamber through a thin Brewster's-angled (0.5 mm thin) fused silica
window which minimizes reflection losses of the broadband incoming
pulses. The crystalline specimens are mounted on a custom-built
multi-axes goniometer (Fig.~\ref{fig:4}(b)), which is used for accurate
positioning of the samples at the beam focus and for precision control
of the crystal orientation with respect to the polarization vector of
the incoming beam.

The custom-built goniometer provides key functions for high harmonic
studies in solids, outperforming commercial x-ray goniometers. It allows
sample translation to undamaged regions while maintaining fixed
three-dimensional rotation axes relative to the laser path, using two
orthogonal translation stages atop rotational stages (Fig.~\ref{fig:4}(b)).
Rotation stage 1 turns the entire assembly around the laser axis, while
stages 2 and 3 enable rotation about two perpendicular axes, ensuring
the sample surface is perpendicular to the laser propagation. A copper
weight mounted on top of the rotating platform counterbalances the
torque produced by the rotation stage 1 during its rotation. The
translation stages provide micrometer-precision access to the entire
sample surface and facilitate sample exchange without any additional
rotational axis adjustments.

An additional linear stage (Fig.~\ref{fig:1}) translates the entire sample holder
unit (goniometer) along the laser propagation axis to correctly position the sample within the focus.
The sample surface can be visually inspected using a 50$\times$ magnification optical system and camera
located outside the HHG chamber (Fig.~\ref{fig:4}(a)).

\subsection{Imaging and spatial filtering assembly}

As alluded to earlier, the high harmonic yield is highly sensitive to
the surface quality of the thin crystal sample. Consequently, any
displacement of the laser focus on the sample surface during high
harmonic measurements (for instance versus crystal angle) can lead to
misleading variations. Therefore, to precisely study high harmonic
anisotropy in crystalline solids, the signal must originate purely from
the crystal's intrinsic anisotropy. In a typical orientation-dependent
yield measurement, a specific crystal axis is rotated on the laser
polarization plane. If the sample's rotation axis does not precisely
coincide with the laser propagation axis, the focal spot of the laser
beam will move across the sample surface and thus generate harmonics
from different parts of the sample. To mitigate this issue the following
strategy has been developed. A crosshair (Fig.~\ref{fig:4}(d)) made of a 40 $\mu$m
thick copper wire is incorporated onto the sample holder and provides a
permanent reference for aligning the laser beam such that its
propagation axis always coincides with the rotation axis of the sample.

Figure \ref{fig:4}(a) illustrates the layout of the f-to-2f imaging system used
for the visualization and measurement of the laser beam profile. The
beam can also illuminate the crosshair in the sample plane, thereby
allowing the crosshair to be visualized indirectly. The optics of the
f-to-2f imaging system include a bare-aluminum-coated toroidal mirror
(f = 10 cm), placed 15 cm away from the sample plane and a biconvex lens
(f = 17.5 cm), placed another 30 cm away from the focus of the toroidal
mirror. A set of three fused silica wedges placed after the toroidal
mirror focus are used to attenuate the laser power before reaching a
high resolution (pixel size 4.6 $\mu$m) beam profiler (Dataray). A nearly
Gaussian beam profile (Fig.~\ref{fig:4}(a)) with a 75 $\mu$m diameter (FWHM) was
typically observed at the focus. Measurements of the beam profile across
varying intensities using the intensity control module showed no
detectable changes, indicating stable and reproducible focusing on the
sample.

To monitor and correct any misalignments between the sample rotation
axis and the laser propagation axis, the steering and focusing mirror mounts
(Fig.~\ref{fig:4}(a)) are iteratively adjusted using the crosshair as a reference,
as shown in Fig.~\ref{fig:4}(f). Once the alignment is complete, the sample is
brought into the beam path. The apparatus can spatially select high
harmonics from a specific focal area by using a ceramic pinhole on a 3D
translation stage at the toroidal mirror's focal plane (Fig.~\ref{fig:4}(a)). When
spatial filtering is not required, the ceramic pinhole is typically
removed from the beam path.

\subsection{Dual-spectrometer}

\begin{figure}[!tb]
\centering
\includegraphics[width=0.95\columnwidth,keepaspectratio,height=1\textheight]{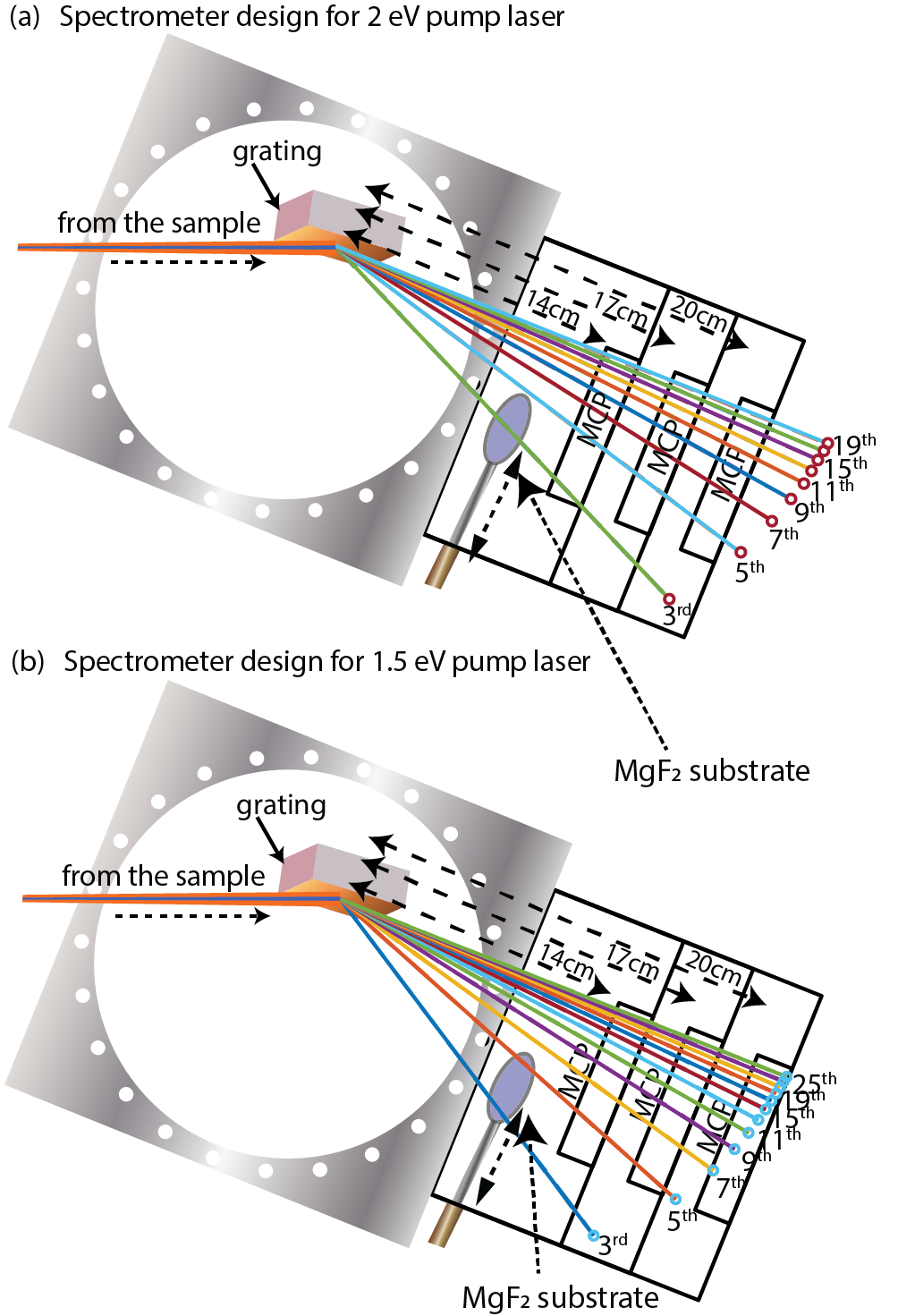}
\caption{EUV spectrometer design. (a), (b) Design of the EUV spectrometer for measuring high harmonics emission driven by visible and NIR driving pulses respectively. The optimal arrangement (imaging of the source and maximal bandwidth) is attained when the MCP detector is placed at a distance of 17 cm from the center of the grating. Harmonic orders (first order diffraction) are denoted by numbers such as 3$^{\mathrm{rd}}$, 5$^{\mathrm{th}}$, 7$^{\mathrm{th}}$ and so on.}
\label{fig:5}
\end{figure}

With the pickup wedge retracted from the beam path, optical and UV-EUV
radiation enter the spectrometer chamber (Fig.~\ref{fig:1}). There, a Hitachi,
aberration-corrected concave grating (001-0266) is used to spectrally
decompose the extreme ultraviolet beam and image the harmonic source on
the detector. To allow a 5$^{\circ}$ grazing incidence of the beam on the
grating, so as to optimize reflection and remain as close as possible to
the angle for which the grating attains it highest imaging performance,
the spectrometer chamber is rotated by $\sim$72.5$^{\circ}$ with respect
to the HHG chamber. Our main detector is a 3-stage microchannel plate
(MCP) with a diameter of 6.8 cm and it is placed at a distance of 17 cm
away from the grating center. To allow operational flexibility and fine
adjustment of the imaging conditions the MCP is placed on an adjustable
vacuum expansion bellow\cite{Kita1983,Harada1999} (Fig.~\ref{fig:5}). In our geometry
the spectrometer can simultaneously record harmonics in the energy range
from $\sim$ 7 eV to 40 eV without any additional adjustment.
To eliminate the second-order diffraction of high harmonics, a 200 $\mu$m
thick magnesium fluoride (MgF\textsubscript{2}) substrate is inserted in
the chamber, and it is positioned close to the front plate of the MCP
detector by an UHV linear motion feedthrough. As MgF$_{2}$ absorbs radiation with energies greater than 10 eV, the substrate
is positioned within the diffracted beam path to allow the transmission of
first-order diffraction at energies \textless{} 10 eV but block harmonics with energies
\textgreater{} 10 eV originating from higher-order diffraction.

The sensitivity of MCP detectors is generally poor for photon energies
below 10 eV. To efficiently detect lower order harmonics a second
spectrometer (McPherson model no.: 234/302) optimized for operation in
the UV/VUV range is integrated into our apparatus. To enable
simultaneous acquisition with both spectrometers, the zeroth-order
diffraction from the Hitachi grating is redirected to the McPherson
spectrometer using a bare aluminum mirror (Fig.~\ref{fig:1}) placed such as not to
obstruct the path of the first order diffraction of the grating to the
MCP. Moreover, to suppress the optical beam, a series of VUV bandpass
filters are employed at the entrance of the McPherson spectrometer. The
filters are attached to externally adjustable vacuum valves which
facilitate their insertion into the beam path. The central transmission
energies of the filters are 4.8 eV and 6.4 eV, with FWHM values of 0.5
eV and 0.8 eV respectively, and are sufficient for capturing the entire
bandwidth of the third harmonic for both driving pulse settings. The
detector of the McPherson spectrometer is comprised of a fiber-coupled
multi-channel plate/phosphor screen and an Andor back illuminated CCD
camera, with high sensitivity in the UV/VUV (3-8 eV) energy range.

\subsection{Data acquisition}

Accurately measuring the yield of high harmonics as a function of
electric field strength or crystal orientation requires long acquisition
times. This is because the process involves mechanically adjusting both
the translation and rotation stages to vary the driving-field amplitude
and the crystal orientation, in addition to the necessary integration
time at each measurement point. In these kinds of measurements, it is
absolutely necessary that the sample remains free of damage in order to
not compromise on the measurement fidelity. Since the degree of
laser-induced damage scales with the total exposure of the specimen to
the laser field\cite{Rosenfeld1999,Lenzner1999,Smalakys2019,Moller2007,Hong2013}, the minimization of the
effective exposure time during data acquisition is crucial to
maintaining the sample quality.

To this end a motorized beam shutter (Fig.~\ref{fig:1}) driven by the acquisition software
is placed before the HHG chamber to limit the exposure of the sample to
the laser beam only for a very short time (typically $\sim$100
milliseconds), within which the cameras of the dual-spectrometer module
record high harmonic spectra while blocking the beam at times when the
sample is rotating or the intensity control unit is adjusting the linear
reflector position to modify the intensity. With this approach, during a
long acquisition of data -- typically requiring several minutes -- the
solid specimen is exposed to the intense laser pulse only for a few
seconds.

A LabVIEW-based, home-made software is used to interface all externally
controlled modules of the experimental apparatus, including the
intensity control, the multi-axis goniometer's rotation and translation
stages, both CCD cameras in the dual-spectrometer, the photodiode used
to track the laser stability (typically better than 2\%), and the beam
shutter. The high degree of automation of critical operations enables
rapid acquisition of experimental data and ensures measurements of an
extensive volume of data under identical experimental conditions.

\section{ELECTRIC FIELD CALIBRATION}

Accurate knowledge of the absolute driving laser field amplitude is
essential for high harmonic spectroscopy because it is an important
parameter in numerous approaches aiming at extracting critical
information about the studied solids\cite{Luu2015,Lakhotia2020,Vampa2017,Huttner2017,Liu2018}.
Ideally an attosecond streaking setup integrated in the apparatus could
serve this goal by providing direct access into the driving field
waveform \cite{Goulielmakis2004,Goulielmakis2008}. Yet, in order to minimize the
complexity of our apparatus and avoid additional alignment burden we
opted here for a different approach in which field calibration is
performed externally.

\begin{figure}[h!]
\centering
\includegraphics[width=0.82\columnwidth,keepaspectratio,height=1\textheight]{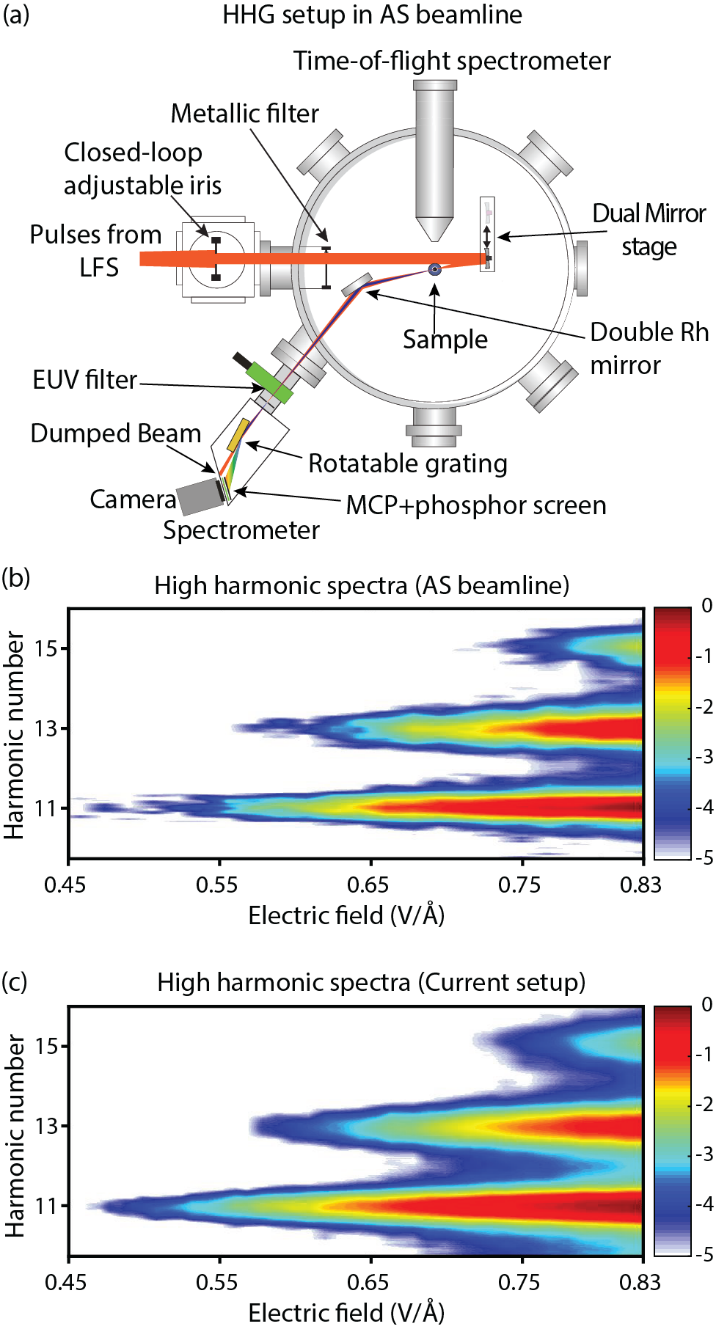}
\caption{Determination of the absolute electric field amplitude and calibration of the apparatus. (a) Schematic of the attosecond (AS) beamline used for the pulse electric field calibration. (b) Recorded high harmonic yield of NIR pulses versus peak electric field determined by attosecond streaking. (c) Calibrated absolute electric field strength-dependent high harmonics from the developed HHG apparatus.}
\label{fig:6}
\end{figure}

To achieve an accurate external calibration of the electric field, we
employed the attosecond (AS) beamline\cite{Schultze2011} in our laboratory.
Using the same laser source for consistency, we first conducted
attosecond streaking measurements to determine the precise peak electric
field amplitude of the laser pulses. Subsequently, we generated high
harmonics in a 150 nm-thick fused silica sample for a range of known
field strengths. The generated high harmonics were then directed into an
EUV spectrometer using a pair of broadband rhodium (Rh) mirrors, as
illustrated in Fig.~\ref{fig:6}(a). Figure \ref{fig:6}(b) displays the measured spectra as a
function of the absolute peak electric field of the driving
near-infrared pulses.

In the next step, the same fused silica sample was transferred to the
newly developed apparatus and driven by the same laser pulses to
generate high harmonics under identical conditions. In this setup,
attosecond streaking is not available, and the electric field cannot be
measured directly; instead, only the laser power is monitored using the
photodiode that also tracks the long-term laser stability. We therefore
recorded the high harmonic spectra as a function of the measured laser
power. To relate these measurements to an absolute electric field scale,
we established a conversion between laser power and field amplitude by
requiring that the onset fields of the higher-order harmonics (orders
\textgreater{} 11) coincide with those obtained in the attosecond
beamline calibration (compare Fig.~\ref{fig:6}(c) and Fig.~\ref{fig:6}(d)). This procedure
ensures a consistent, absolute field calibration across both measurement
platforms.

\section{Representative HHG measurements}

To demonstrate the capabilities and quantitative reliability of the
apparatus, we performed a series of representative high harmonic
generation measurements using both amorphous and crystalline solids.
These measurements serve two complementary purposes. First, they
validate that the apparatus can stably generate and detect high-order
harmonics over broad spectral ranges and field strengths. Second, they
highlight the level of precision attainable in field- and
angle-dependent studies -- an essential requirement for extracting
material properties and benchmarking theoretical models.

\begin{figure*}
\centering
\includegraphics[width=1.8\columnwidth,keepaspectratio,height=1\textheight]{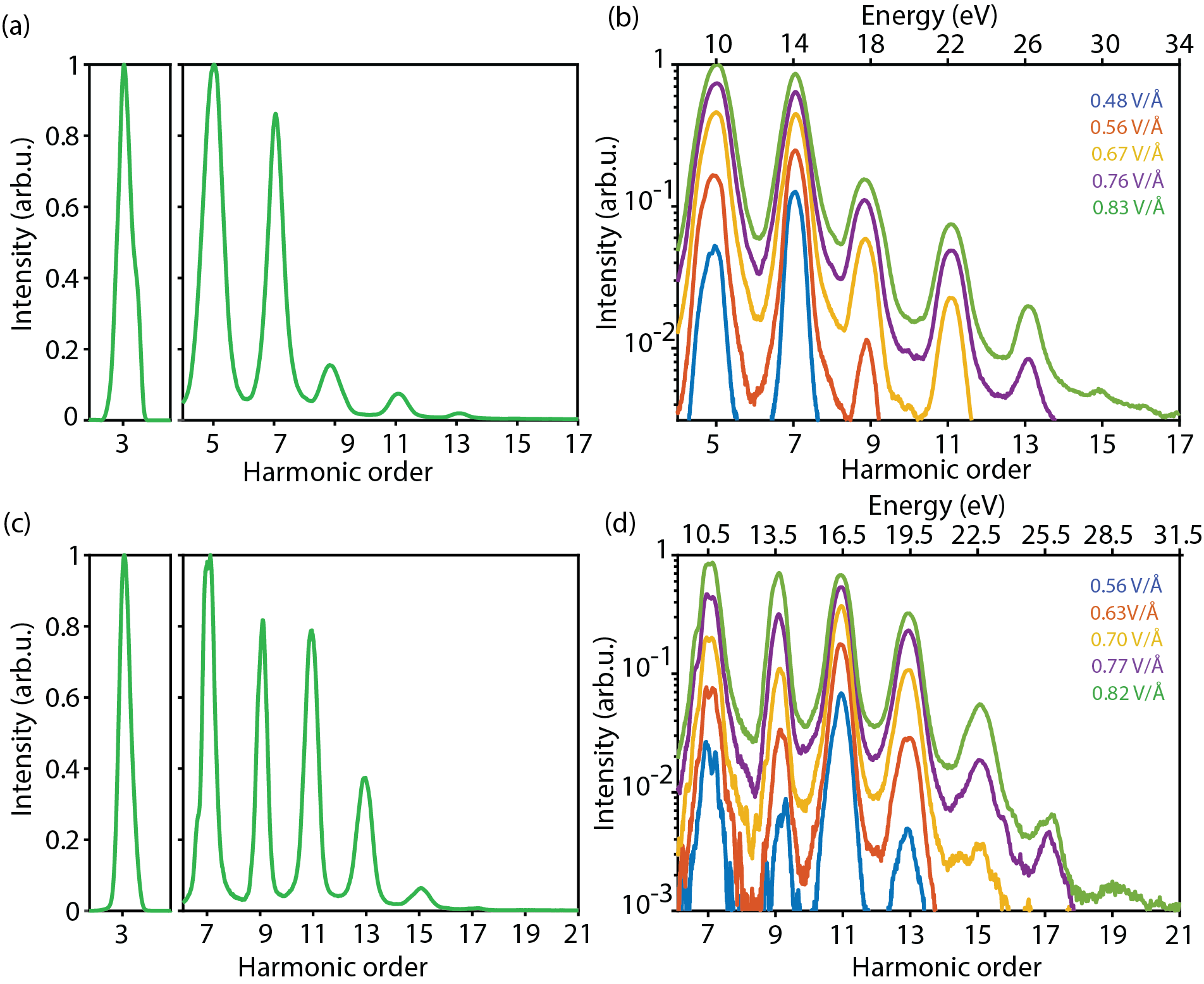}
\caption{High harmonic generation spectra in a fused silica nanofilm (150 nm). Using visible laser pulse (2 eV) (a) and a NIR laser pulse (1.5 eV) (c). (b), (d) High harmonics spectra for the visible and NIR laser pulses respectively at different electric field strengths.}
\label{fig:7}
\end{figure*}

We begin with HHG from a free-standing, amorphous 150-nm fused silica
membrane excited by visible laser pulses (2.0 eV). Figure \ref{fig:7}(a) shows a
representative harmonic spectrum recorded at a peak driving field of
0.83 V/Å, spanning harmonic orders from the 3\textsuperscript{rd} to the
13\textsuperscript{th}. As the electric field strength is increased from
0.48 V/Å to 0.83 V/Å (Fig.~\ref{fig:7}(b)), higher harmonics up to the
15\textsuperscript{th} order ($\approx$ 30 eV) appear, consistent with the
expected field-dependent extension of the cutoff. The equivalent
measurements using NIR pulses at 1.5 eV (Figs. \ref{fig:7}(c)--(d)) yield cutoffs
reaching the 19\textsuperscript{th} harmonic for fields approaching 0.82
V/Å. The smooth evolution of the harmonic cutoff and the absence of
fluctuations across field steps reflect the stability of the intensity
control module and the high repeatability of the beam delivery and
detection systems.

In order to explore the capability of the apparatus to perform precision
measurements of the anisotropy of the high harmonics yield versus
crystal angle, we used a single crystal, $\sim$10 $\mu$m thick,
magnesium oxide -- MgO (100) sample. MgO is selected as a benchmark
material because of its high damage threshold, and clear symmetry-driven
anisotropy, which make it ideally suited for validating apparatus
precision.

\begin{SCfigure*}
\centering
\fbox{\includegraphics[width=1.5\columnwidth,keepaspectratio,height=1\textheight]{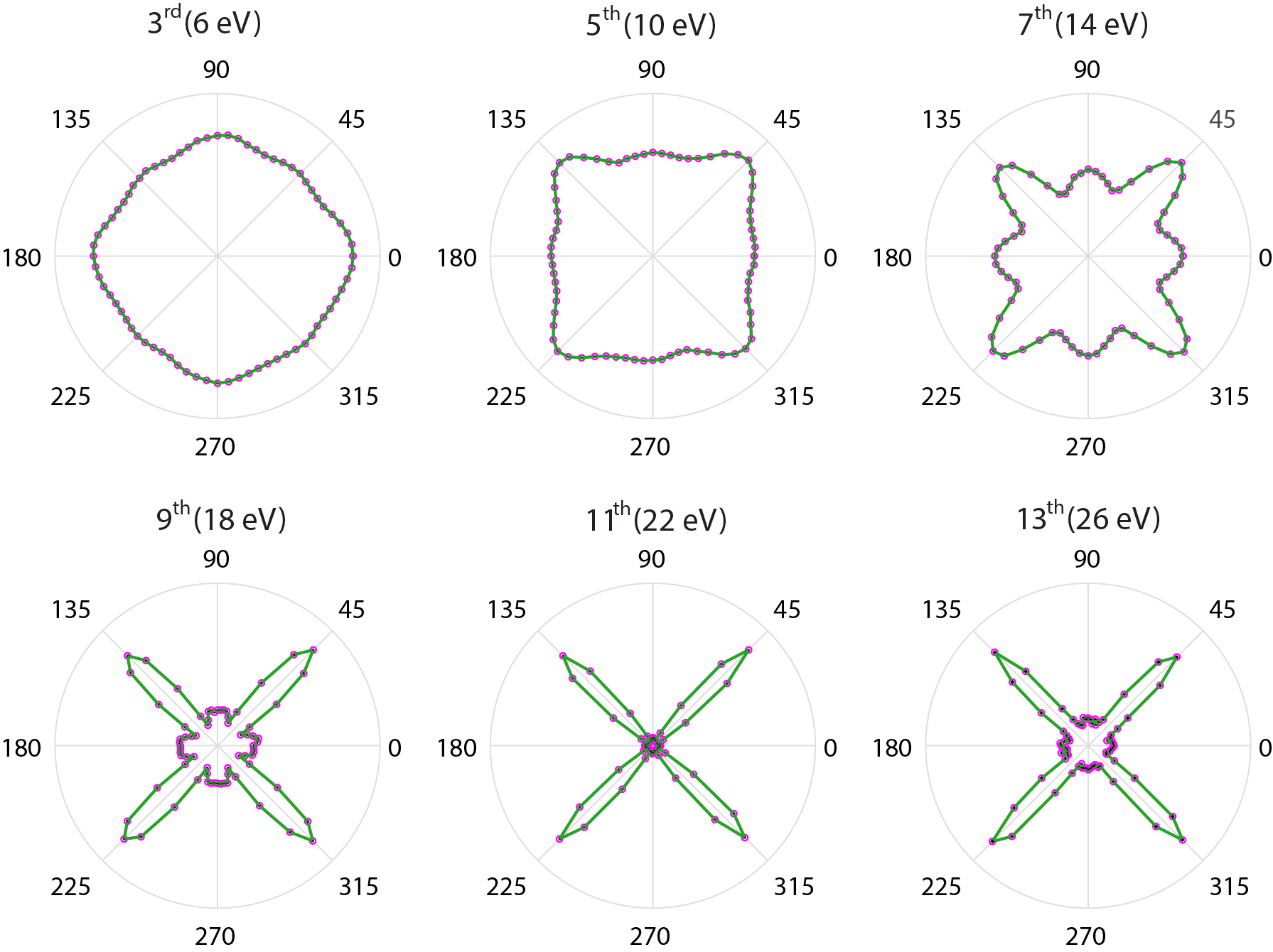}}
\caption{High harmonic yield versus crystal angle in MgO (100) generated by linearly polarized laser pulses centered at an energy of 2 eV. The radial direction represents the intensity (in arbitrary units) whereas the azimuthal direction represents the angle of rotation of the crystal with respect to the polarization axis of the linearly polarized laser. The magenta circles show experimental data, with the green line serving as a guide to the eye.}
\label{fig:8}
\end{SCfigure*}

The sample is shone by the linearly-polarized visible laser pulses, and
the harmonic yield is recorded as a function of the crystal angle, with
the laser polarization parallel to the MgO (100) plane. Here, the
crystal angle is defined as the angle between the laser polarization
vector and the lattice vector {[}001{]} (or equivalently {[}010{]}, due
to MgO's 90$^\circ$ rotational symmetry) on the MgO (100) plane. Figure \ref{fig:8}
illustrates the recorded harmonic yields (integrated over entire
spectral width of each harmonic spectral peak) over the range from
0\textsuperscript{o} to 360\textsuperscript{o} at an angular step of
5\textsuperscript{o}. The yield of the recorded harmonics exhibits a
90\textsuperscript{o} periodicity, compatible with the crystal structure
of MgO\cite{You2017,Hussain2020,Heinrich2021,Korobenko2021}. Despite a minimal number of
acquisitions (3) per angle the high symmetry of the data for all
harmonics recorded is indicative for the quality of the measurements.

\sidecaptionvpos{figure}{c}

\sidecaptionvpos{figure}{c}
\begin{SCfigure*}
\centering
\fbox{\includegraphics[width=1.5\columnwidth,keepaspectratio,height=1\textheight]{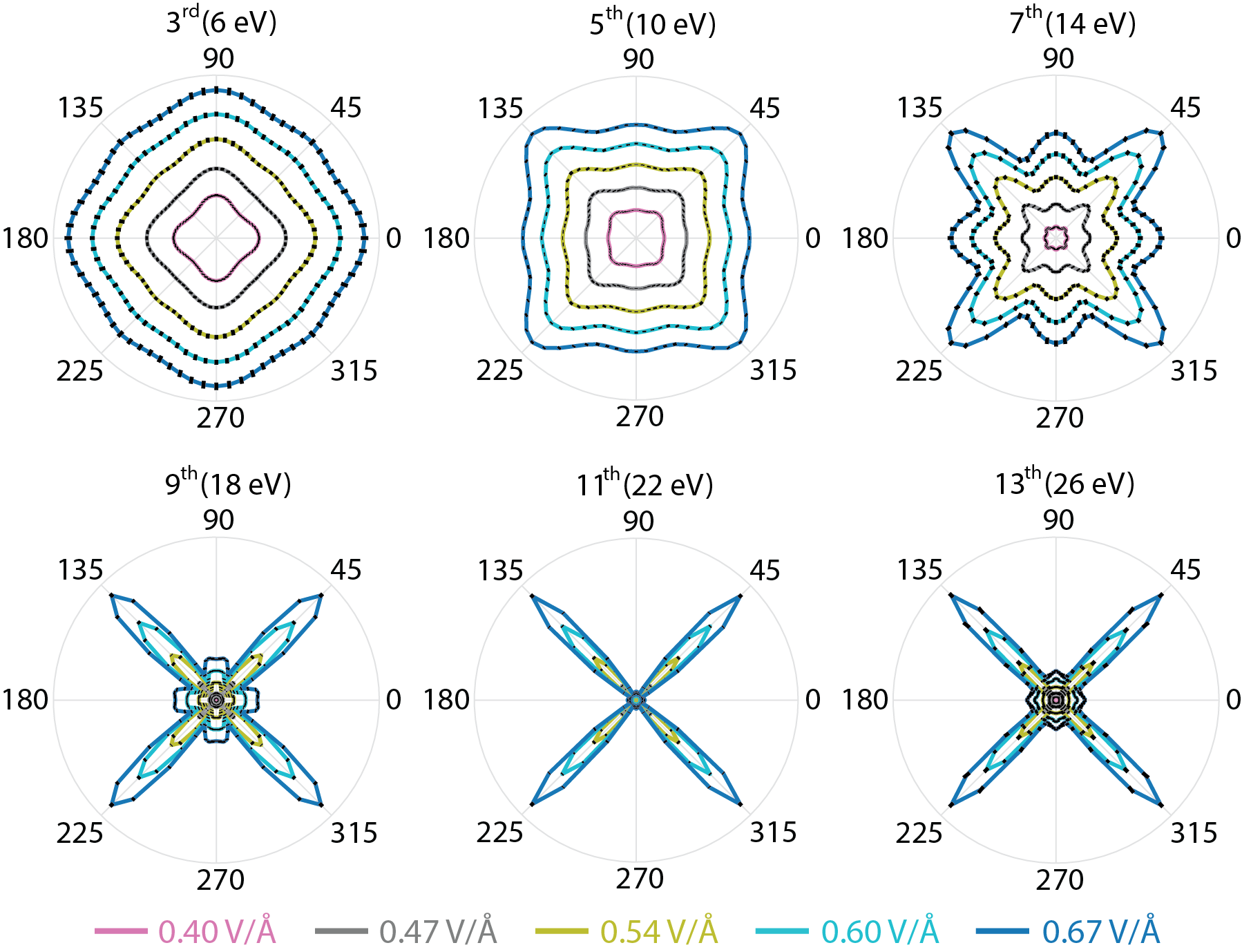}}
\caption{High harmonic yields in MgO as a function of crystal rotation about an axis perpendicular to the sample surface, and electric field strength. The high harmonics are measured for five representative field strengths of 0.40 V/Å (magenta), 0.47 V/Å (grey), 0.54 V/Å (olive), 0.60 V/Å (cyan) and 0.67 V/Å (blue). The data are shown in black (5° step) while the width of each data segment represent the standard error. Solid lines serve as guides to the eye. Each harmonic signal is normalized with respect to its maximum yield to enhance visibility.}
\label{fig:9}
\end{SCfigure*}

Figure \ref{fig:9} shows the anisotropy of the high harmonic yield in crystalline
MgO (100) plane for a range of driving field amplitudes 0.40, 0.47,
0.54, 0.60 and 0.67 V/Å. The high precision of the measurements is best
highlighted by the evaluation of the standard errors (black segments).
The standard error is evaluated by performing 3 consecutive measurements
of the yield of the harmonics at each angle (integration time
$\sim$100 ms) as well as by averaging over the \emph{n} (8
for MgO) symmetric segments of the angular variation of the harmonic
yield. Between rotation steps, the motorized shutter blocks the laser
beam for 2 s, giving a total scan time of about 3 min for a full
rotational scan per intensity setting. Across all field strengths and
angles, the error bars are exceptionally small -- typically only a few
percent of the total signal -- while the angular modulation spans tens of
percent. This separation of scales demonstrates that the observed
anisotropy is governed by the intrinsic crystal response, not by
instrumental noise or drift.~The small statistical uncertainties further
confirm that the goniometer alignment, beam-pointing stability, and
detector performance remain constant throughout extended measurement
sequences, enabling quantitative comparison of angle-dependent yields
across different field strengths.

Taken together, the measurements in Figs. \ref{fig:7}--\ref{fig:9} provide a comprehensive
validation of the apparatus. The fused silica results confirm stable and
broadband HHG detection under varying field strengths, while the MgO
measurements reveal the exceptional angular and statistical precision
required for quantitative anisotropy studies. The combination of
negligible error bars, symmetry-clean angular traces, and consistent
field-dependent trends establishes that the apparatus delivers
drift-free, reproducible, and quantitative HHG measurements in bulk
solids. These capabilities form a robust foundation for advanced
spectroscopic applications, including band-structure mapping,
symmetry-resolved strong-field studies, and the reconstruction of
valence-electron dynamics under tailored light fields.

\section{CONCLUSION AND OUTLOOK}

We have developed a fully integrated apparatus designed for precision
high harmonic spectroscopy in bulk solids, combining dispersion-free
intensity control of few-cycle driving fields, sub-micrometer and
sub-degree sample positioning, high-fidelity imaging and spatial
filtering of the HHG source, and synchronized UV/VUV--EUV detection.
Through absolute electric-field calibration via attosecond streaking and
careful suppression of geometrical and alignment artefacts, the system
enables reproducible, quantitative measurements of field- and
angle-dependent harmonic emission across a broad spectral range.
Representative results from amorphous and crystalline solids demonstrate
the apparatus's ability to capture subtle anisotropies and
field-strength-dependent cutoffs with high accuracy. Beyond the
examples presented here, the modular design of the platform allows
adaptation to diverse material systems, driving wavelengths, and
advanced spectroscopic methodologies. As such, the apparatus provides a
robust foundation for future efforts aimed at reconstructing
valence-electron potentials, mapping band structures with high
precision, and exploring strong-field phenomena in emerging quantum
materials.~\\

\begin{acknowledgments}
This work was jointly supported by the European Research Council (project number - 101098243) and Deutsche Forschungsgemeinschaft (project numbers - 441234705 and 437567992).
\end{acknowledgments}

\section*{Author Declarations}

\subsection*{Conflict of Interest}
The authors declare that they have no competing interests.

\subsection*{Author Contributions}
SM, ZP, HYK, MZ and PK designed and developed the apparatus. SM and PK performed the experiments. SM, PK and EG prepared the manuscript. EG conceived and supervised the experiments.

\section*{Data Availability}
The data that support the findings of this study are available from the corresponding author upon reasonable request.

\section*{References}
\nocite{*}
\bibliography{mainNotes}

\begin{thebibliography}{37}%
\makeatletter
\providecommand \@ifxundefined [1]{%
 \@ifx{#1\undefined}
}%
\providecommand \@ifnum [1]{%
 \ifnum #1\expandafter \@firstoftwo
 \else \expandafter \@secondoftwo
 \fi
}%
\providecommand \@ifx [1]{%
 \ifx #1\expandafter \@firstoftwo
 \else \expandafter \@secondoftwo
 \fi
}%
\providecommand \natexlab [1]{#1}%
\providecommand \enquote  [1]{``#1''}%
\providecommand \bibnamefont  [1]{#1}%
\providecommand \bibfnamefont [1]{#1}%
\providecommand \citenamefont [1]{#1}%
\providecommand \href@noop [0]{\@secondoftwo}%
\providecommand \href [0]{\begingroup \@sanitize@url \@href}%
\providecommand \@href[1]{\@@startlink{#1}\@@href}%
\providecommand \@@href[1]{\endgroup#1\@@endlink}%
\providecommand \@sanitize@url [0]{\catcode `\\12\catcode `\$12\catcode `\&12\catcode `\#12\catcode `\^12\catcode `\_12\catcode `\%12\relax}%
\providecommand \@@startlink[1]{}%
\providecommand \@@endlink[0]{}%
\providecommand \url  [0]{\begingroup\@sanitize@url \@url }%
\providecommand \@url [1]{\endgroup\@href {#1}{\urlprefix }}%
\providecommand \urlprefix  [0]{URL }%
\providecommand \Eprint [0]{\href }%
\providecommand \doibase [0]{https://doi.org/}%
\providecommand \selectlanguage [0]{\@gobble}%
\providecommand \bibinfo  [0]{\@secondoftwo}%
\providecommand \bibfield  [0]{\@secondoftwo}%
\providecommand \translation [1]{[#1]}%
\providecommand \BibitemOpen [0]{}%
\providecommand \bibitemStop [0]{}%
\providecommand \bibitemNoStop [0]{.\EOS\space}%
\providecommand \EOS [0]{\spacefactor3000\relax}%
\providecommand \BibitemShut  [1]{\csname bibitem#1\endcsname}%
\let\auto@bib@innerbib\@empty
\bibitem [{\citenamefont {Ghimire}\ \emph {et~al.}(2011)\citenamefont {Ghimire}, \citenamefont {DiChiara}, \citenamefont {Sistrunk}, \citenamefont {Agostini}, \citenamefont {DiMauro},\ and\ \citenamefont {Reis}}]{Ghimire2011}%
  \BibitemOpen
  \bibfield  {author} {\bibinfo {author} {\bibfnamefont {S.}~\bibnamefont {Ghimire}}, \bibinfo {author} {\bibfnamefont {A.~D.}\ \bibnamefont {DiChiara}}, \bibinfo {author} {\bibfnamefont {E.}~\bibnamefont {Sistrunk}}, \bibinfo {author} {\bibfnamefont {P.}~\bibnamefont {Agostini}}, \bibinfo {author} {\bibfnamefont {L.~F.}\ \bibnamefont {DiMauro}},\ and\ \bibinfo {author} {\bibfnamefont {D.~A.}\ \bibnamefont {Reis}},\ }\bibfield  {title} {\enquote {\bibinfo {title} {Observation of high-order harmonic generation in a bulk crystal},}\ }\href {https://doi.org/10.1038/nphys1847} {\bibfield  {journal} {\bibinfo  {journal} {Nature Physics}\ }\textbf {\bibinfo {volume} {7}},\ \bibinfo {pages} {138--141} (\bibinfo {year} {2011})}\BibitemShut {NoStop}%
\bibitem [{\citenamefont {Luu}\ \emph {et~al.}(2015)\citenamefont {Luu}, \citenamefont {Garg}, \citenamefont {Kruchinin}, \citenamefont {Moulet}, \citenamefont {Hassan},\ and\ \citenamefont {Goulielmakis}}]{Luu2015}%
  \BibitemOpen
  \bibfield  {author} {\bibinfo {author} {\bibfnamefont {T.~T.}\ \bibnamefont {Luu}}, \bibinfo {author} {\bibfnamefont {M.}~\bibnamefont {Garg}}, \bibinfo {author} {\bibfnamefont {S.~Y.}\ \bibnamefont {Kruchinin}}, \bibinfo {author} {\bibfnamefont {A.}~\bibnamefont {Moulet}}, \bibinfo {author} {\bibfnamefont {M.~T.}\ \bibnamefont {Hassan}},\ and\ \bibinfo {author} {\bibfnamefont {E.}~\bibnamefont {Goulielmakis}},\ }\bibfield  {title} {\enquote {\bibinfo {title} {Extreme ultraviolet high-harmonic spectroscopy of solids},}\ }\href {https://doi.org/10.1038/nature14456} {\bibfield  {journal} {\bibinfo  {journal} {Nature}\ }\textbf {\bibinfo {volume} {521}},\ \bibinfo {pages} {498--502} (\bibinfo {year} {2015})}\BibitemShut {NoStop}%
\bibitem [{\citenamefont {Garg}\ \emph {et~al.}(2016)\citenamefont {Garg}, \citenamefont {Zhan}, \citenamefont {Luu}, \citenamefont {Lakhotia}, \citenamefont {Klostermann}, \citenamefont {Guggenmos},\ and\ \citenamefont {Goulielmakis}}]{Garg2016}%
  \BibitemOpen
  \bibfield  {author} {\bibinfo {author} {\bibfnamefont {M.}~\bibnamefont {Garg}}, \bibinfo {author} {\bibfnamefont {M.}~\bibnamefont {Zhan}}, \bibinfo {author} {\bibfnamefont {T.~T.}\ \bibnamefont {Luu}}, \bibinfo {author} {\bibfnamefont {H.}~\bibnamefont {Lakhotia}}, \bibinfo {author} {\bibfnamefont {T.}~\bibnamefont {Klostermann}}, \bibinfo {author} {\bibfnamefont {A.}~\bibnamefont {Guggenmos}},\ and\ \bibinfo {author} {\bibfnamefont {E.}~\bibnamefont {Goulielmakis}},\ }\bibfield  {title} {\enquote {\bibinfo {title} {Multi-petahertz electronic metrology},}\ }\href {https://doi.org/10.1038/nature19821} {\bibfield  {journal} {\bibinfo  {journal} {Nature}\ }\textbf {\bibinfo {volume} {538}},\ \bibinfo {pages} {359--363} (\bibinfo {year} {2016})}\BibitemShut {NoStop}%
\bibitem [{\citenamefont {J\"urgens}\ \emph {et~al.}(2020)\citenamefont {J\"urgens}, \citenamefont {Liewehr}, \citenamefont {Kruse}, \citenamefont {Peltz}, \citenamefont {Engel}, \citenamefont {Husakou}, \citenamefont {Witting}, \citenamefont {Ivanov}, \citenamefont {Vrakking}, \citenamefont {Fennel},\ and\ \citenamefont {Mermillod-Blondin}}]{Jurgens2020}%
  \BibitemOpen
  \bibfield  {author} {\bibinfo {author} {\bibfnamefont {P.}~\bibnamefont {J\"urgens}}, \bibinfo {author} {\bibfnamefont {B.}~\bibnamefont {Liewehr}}, \bibinfo {author} {\bibfnamefont {B.}~\bibnamefont {Kruse}}, \bibinfo {author} {\bibfnamefont {C.}~\bibnamefont {Peltz}}, \bibinfo {author} {\bibfnamefont {D.}~\bibnamefont {Engel}}, \bibinfo {author} {\bibfnamefont {A.}~\bibnamefont {Husakou}}, \bibinfo {author} {\bibfnamefont {T.}~\bibnamefont {Witting}}, \bibinfo {author} {\bibfnamefont {M.}~\bibnamefont {Ivanov}}, \bibinfo {author} {\bibfnamefont {M.~J.~J.}\ \bibnamefont {Vrakking}}, \bibinfo {author} {\bibfnamefont {T.}~\bibnamefont {Fennel}},\ and\ \bibinfo {author} {\bibfnamefont {A.}~\bibnamefont {Mermillod-Blondin}},\ }\bibfield  {title} {\enquote {\bibinfo {title} {Origin of strong-field-induced low-order harmonic generation in amorphous quartz},}\ }\href {https://doi.org/10.1038/s41567-020-0943-4} {\bibfield  {journal} {\bibinfo  {journal} {Nature Physics}\ }\textbf {\bibinfo {volume} {16}},\
  \bibinfo {pages} {1035--1039} (\bibinfo {year} {2020})}\BibitemShut {NoStop}%
\bibitem [{\citenamefont {Liu}\ \emph {et~al.}(2017)\citenamefont {Liu}, \citenamefont {Li}, \citenamefont {You}, \citenamefont {Ghimire}, \citenamefont {Heinz},\ and\ \citenamefont {Reis}}]{Liu2017}%
  \BibitemOpen
  \bibfield  {author} {\bibinfo {author} {\bibfnamefont {H.}~\bibnamefont {Liu}}, \bibinfo {author} {\bibfnamefont {Y.}~\bibnamefont {Li}}, \bibinfo {author} {\bibfnamefont {Y.~S.}\ \bibnamefont {You}}, \bibinfo {author} {\bibfnamefont {S.}~\bibnamefont {Ghimire}}, \bibinfo {author} {\bibfnamefont {T.~F.}\ \bibnamefont {Heinz}},\ and\ \bibinfo {author} {\bibfnamefont {D.~A.}\ \bibnamefont {Reis}},\ }\bibfield  {title} {\enquote {\bibinfo {title} {High-harmonic generation from an atomically thin semiconductor},}\ }\href {https://doi.org/10.1038/nphys3946} {\bibfield  {journal} {\bibinfo  {journal} {Nature Physics}\ }\textbf {\bibinfo {volume} {13}},\ \bibinfo {pages} {262--265} (\bibinfo {year} {2017})}\BibitemShut {NoStop}%
\bibitem [{\citenamefont {Yoshikawa}, \citenamefont {Tamaya},\ and\ \citenamefont {Tanaka}(2017)}]{Yoshikawa2017}%
  \BibitemOpen
  \bibfield  {author} {\bibinfo {author} {\bibfnamefont {N.}~\bibnamefont {Yoshikawa}}, \bibinfo {author} {\bibfnamefont {T.}~\bibnamefont {Tamaya}},\ and\ \bibinfo {author} {\bibfnamefont {K.}~\bibnamefont {Tanaka}},\ }\bibfield  {title} {\enquote {\bibinfo {title} {High-harmonic generation in graphene enhanced by elliptically polarized light excitation},}\ }\href {https://doi.org/10.1126/science.aam8861} {\bibfield  {journal} {\bibinfo  {journal} {Science}\ }\textbf {\bibinfo {volume} {356}},\ \bibinfo {pages} {736--738} (\bibinfo {year} {2017})}\BibitemShut {NoStop}%
\bibitem [{\citenamefont {Silva}\ \emph {et~al.}(2018)\citenamefont {Silva}, \citenamefont {Blinov}, \citenamefont {Rubtsov}, \citenamefont {Smirnova},\ and\ \citenamefont {Ivanov}}]{Silva2018}%
  \BibitemOpen
  \bibfield  {author} {\bibinfo {author} {\bibfnamefont {R.~E.~F.}\ \bibnamefont {Silva}}, \bibinfo {author} {\bibfnamefont {I.~V.}\ \bibnamefont {Blinov}}, \bibinfo {author} {\bibfnamefont {A.~N.}\ \bibnamefont {Rubtsov}}, \bibinfo {author} {\bibfnamefont {O.}~\bibnamefont {Smirnova}},\ and\ \bibinfo {author} {\bibfnamefont {M.}~\bibnamefont {Ivanov}},\ }\bibfield  {title} {\enquote {\bibinfo {title} {High-harmonic spectroscopy of ultrafast many-body dynamics in strongly correlated systems},}\ }\href {https://doi.org/10.1038/s41566-018-0129-2} {\bibfield  {journal} {\bibinfo  {journal} {Nature Photonics}\ }\textbf {\bibinfo {volume} {12}},\ \bibinfo {pages} {266--270} (\bibinfo {year} {2018})}\BibitemShut {NoStop}%
\bibitem [{\citenamefont {Murakami}, \citenamefont {Eckstein},\ and\ \citenamefont {Werner}(2018)}]{Murakami2018}%
  \BibitemOpen
  \bibfield  {author} {\bibinfo {author} {\bibfnamefont {Y.}~\bibnamefont {Murakami}}, \bibinfo {author} {\bibfnamefont {M.}~\bibnamefont {Eckstein}},\ and\ \bibinfo {author} {\bibfnamefont {P.}~\bibnamefont {Werner}},\ }\bibfield  {title} {\enquote {\bibinfo {title} {High-harmonic generation in mott insulators},}\ }\href {https://doi.org/10.1103/PhysRevLett.121.057405} {\bibfield  {journal} {\bibinfo  {journal} {Physical Review Letters}\ }\textbf {\bibinfo {volume} {121}},\ \bibinfo {pages} {057405} (\bibinfo {year} {2018})}\BibitemShut {NoStop}%
\bibitem [{\citenamefont {Bauer}\ and\ \citenamefont {Hansen}(2018)}]{Bauer2018}%
  \BibitemOpen
  \bibfield  {author} {\bibinfo {author} {\bibfnamefont {D.}~\bibnamefont {Bauer}}\ and\ \bibinfo {author} {\bibfnamefont {K.~K.}\ \bibnamefont {Hansen}},\ }\bibfield  {title} {\enquote {\bibinfo {title} {High-harmonic generation in solids with and without topological edge states},}\ }\href {https://doi.org/10.1103/PhysRevLett.120.177401} {\bibfield  {journal} {\bibinfo  {journal} {Physical Review Letters}\ }\textbf {\bibinfo {volume} {120}},\ \bibinfo {pages} {177401} (\bibinfo {year} {2018})}\BibitemShut {NoStop}%
\bibitem [{\citenamefont {Baykusheva}\ \emph {et~al.}(2021)\citenamefont {Baykusheva}, \citenamefont {Chac\'on}, \citenamefont {Lu}, \citenamefont {Bailey}, \citenamefont {Sobota}, \citenamefont {Soifer}, \citenamefont {Kirchmann}, \citenamefont {Rotundu}, \citenamefont {Uher}, \citenamefont {Heinz}, \citenamefont {Reis},\ and\ \citenamefont {Ghimire}}]{Baykusheva2021}%
  \BibitemOpen
  \bibfield  {author} {\bibinfo {author} {\bibfnamefont {D.~R.}\ \bibnamefont {Baykusheva}}, \bibinfo {author} {\bibfnamefont {A.}~\bibnamefont {Chac\'on}}, \bibinfo {author} {\bibfnamefont {J.}~\bibnamefont {Lu}}, \bibinfo {author} {\bibfnamefont {T.~P.}\ \bibnamefont {Bailey}}, \bibinfo {author} {\bibfnamefont {J.~A.}\ \bibnamefont {Sobota}}, \bibinfo {author} {\bibfnamefont {H.}~\bibnamefont {Soifer}}, \bibinfo {author} {\bibfnamefont {P.~S.}\ \bibnamefont {Kirchmann}}, \bibinfo {author} {\bibfnamefont {C.}~\bibnamefont {Rotundu}}, \bibinfo {author} {\bibfnamefont {C.}~\bibnamefont {Uher}}, \bibinfo {author} {\bibfnamefont {T.~F.}\ \bibnamefont {Heinz}}, \bibinfo {author} {\bibfnamefont {D.~A.}\ \bibnamefont {Reis}},\ and\ \bibinfo {author} {\bibfnamefont {S.}~\bibnamefont {Ghimire}},\ }\bibfield  {title} {\enquote {\bibinfo {title} {All-optical probe of three-dimensional topological insulators based on high-harmonic generation by circularly polarized laser fields},}\ }\href
  {https://doi.org/10.1021/acs.nanolett.1c02145} {\bibfield  {journal} {\bibinfo  {journal} {Nano Letters}\ }\textbf {\bibinfo {volume} {21}},\ \bibinfo {pages} {8970--8978} (\bibinfo {year} {2021})}\BibitemShut {NoStop}%
\bibitem [{\citenamefont {You}, \citenamefont {Reis},\ and\ \citenamefont {Ghimire}(2017)}]{You2017}%
  \BibitemOpen
  \bibfield  {author} {\bibinfo {author} {\bibfnamefont {Y.~S.}\ \bibnamefont {You}}, \bibinfo {author} {\bibfnamefont {D.~A.}\ \bibnamefont {Reis}},\ and\ \bibinfo {author} {\bibfnamefont {S.}~\bibnamefont {Ghimire}},\ }\bibfield  {title} {\enquote {\bibinfo {title} {Anisotropic high-harmonic generation in bulk crystals},}\ }\href {https://doi.org/10.1038/nphys3955} {\bibfield  {journal} {\bibinfo  {journal} {Nature Physics}\ }\textbf {\bibinfo {volume} {13}},\ \bibinfo {pages} {345--349} (\bibinfo {year} {2017})}\BibitemShut {NoStop}%
\bibitem [{\citenamefont {Lanin}\ \emph {et~al.}(2017)\citenamefont {Lanin}, \citenamefont {Stepanov}, \citenamefont {Fedotov},\ and\ \citenamefont {Zheltikov}}]{Lanin2017}%
  \BibitemOpen
  \bibfield  {author} {\bibinfo {author} {\bibfnamefont {A.~A.}\ \bibnamefont {Lanin}}, \bibinfo {author} {\bibfnamefont {E.~A.}\ \bibnamefont {Stepanov}}, \bibinfo {author} {\bibfnamefont {A.~B.}\ \bibnamefont {Fedotov}},\ and\ \bibinfo {author} {\bibfnamefont {A.~M.}\ \bibnamefont {Zheltikov}},\ }\bibfield  {title} {\enquote {\bibinfo {title} {Mapping the electron band structure by intraband high-harmonic generation in solids},}\ }\href {https://doi.org/10.1364/OPTICA.4.000516} {\bibfield  {journal} {\bibinfo  {journal} {Optica}\ }\textbf {\bibinfo {volume} {4}},\ \bibinfo {pages} {516} (\bibinfo {year} {2017})}\BibitemShut {NoStop}%
\bibitem [{\citenamefont {Luu}\ and\ \citenamefont {W\"orner}(2018)}]{Luu2018}%
  \BibitemOpen
  \bibfield  {author} {\bibinfo {author} {\bibfnamefont {T.~T.}\ \bibnamefont {Luu}}\ and\ \bibinfo {author} {\bibfnamefont {H.~J.}\ \bibnamefont {W\"orner}},\ }\bibfield  {title} {\enquote {\bibinfo {title} {Measurement of the berry curvature of solids using high-harmonic spectroscopy},}\ }\href {https://doi.org/10.1038/s41467-018-03397-4} {\bibfield  {journal} {\bibinfo  {journal} {Nature Communications}\ }\textbf {\bibinfo {volume} {9}},\ \bibinfo {pages} {916} (\bibinfo {year} {2018})}\BibitemShut {NoStop}%
\bibitem [{\citenamefont {Lakhotia}\ \emph {et~al.}(2020)\citenamefont {Lakhotia}, \citenamefont {Kim}, \citenamefont {Zhan}, \citenamefont {Hu}, \citenamefont {Meng},\ and\ \citenamefont {Goulielmakis}}]{Lakhotia2020}%
  \BibitemOpen
  \bibfield  {author} {\bibinfo {author} {\bibfnamefont {H.}~\bibnamefont {Lakhotia}}, \bibinfo {author} {\bibfnamefont {H.~Y.}\ \bibnamefont {Kim}}, \bibinfo {author} {\bibfnamefont {M.}~\bibnamefont {Zhan}}, \bibinfo {author} {\bibfnamefont {S.}~\bibnamefont {Hu}}, \bibinfo {author} {\bibfnamefont {S.}~\bibnamefont {Meng}},\ and\ \bibinfo {author} {\bibfnamefont {E.}~\bibnamefont {Goulielmakis}},\ }\bibfield  {title} {\enquote {\bibinfo {title} {Laser picoscopy of valence electrons in solids},}\ }\href {https://doi.org/10.1038/s41586-020-2440-4} {\bibfield  {journal} {\bibinfo  {journal} {Nature}\ }\textbf {\bibinfo {volume} {583}},\ \bibinfo {pages} {55--59} (\bibinfo {year} {2020})}\BibitemShut {NoStop}%
\bibitem [{\citenamefont {You}\ \emph {et~al.}(2018)\citenamefont {You}, \citenamefont {Cunningham}, \citenamefont {Reis},\ and\ \citenamefont {Ghimire}}]{You2018}%
  \BibitemOpen
  \bibfield  {author} {\bibinfo {author} {\bibfnamefont {Y.~S.}\ \bibnamefont {You}}, \bibinfo {author} {\bibfnamefont {E.}~\bibnamefont {Cunningham}}, \bibinfo {author} {\bibfnamefont {D.~A.}\ \bibnamefont {Reis}},\ and\ \bibinfo {author} {\bibfnamefont {S.}~\bibnamefont {Ghimire}},\ }\bibfield  {title} {\enquote {\bibinfo {title} {Probing periodic potential of crystals via strong-field re-scattering},}\ }\href {https://doi.org/10.1088/1361-6455/aac048} {\bibfield  {journal} {\bibinfo  {journal} {Journal of Physics B: Atomic, Molecular and Optical Physics}\ }\textbf {\bibinfo {volume} {51}},\ \bibinfo {pages} {114002} (\bibinfo {year} {2018})}\BibitemShut {NoStop}%
\bibitem [{\citenamefont {Goulielmakis}\ \emph {et~al.}(2004)\citenamefont {Goulielmakis}, \citenamefont {Uiberacker}, \citenamefont {Kienberger}, \citenamefont {Baltuska}, \citenamefont {Yakovlev}, \citenamefont {Scrinzi}, \citenamefont {Westerwalbesloh}, \citenamefont {Kleineberg}, \citenamefont {Heinzmann}, \citenamefont {Drescher},\ and\ \citenamefont {Krausz}}]{Goulielmakis2004}%
  \BibitemOpen
  \bibfield  {author} {\bibinfo {author} {\bibfnamefont {E.}~\bibnamefont {Goulielmakis}}, \bibinfo {author} {\bibfnamefont {M.}~\bibnamefont {Uiberacker}}, \bibinfo {author} {\bibfnamefont {R.}~\bibnamefont {Kienberger}}, \bibinfo {author} {\bibfnamefont {A.}~\bibnamefont {Baltuska}}, \bibinfo {author} {\bibfnamefont {V.}~\bibnamefont {Yakovlev}}, \bibinfo {author} {\bibfnamefont {A.}~\bibnamefont {Scrinzi}}, \bibinfo {author} {\bibfnamefont {T.}~\bibnamefont {Westerwalbesloh}}, \bibinfo {author} {\bibfnamefont {U.}~\bibnamefont {Kleineberg}}, \bibinfo {author} {\bibfnamefont {U.}~\bibnamefont {Heinzmann}}, \bibinfo {author} {\bibfnamefont {M.}~\bibnamefont {Drescher}},\ and\ \bibinfo {author} {\bibfnamefont {F.}~\bibnamefont {Krausz}},\ }\bibfield  {title} {\enquote {\bibinfo {title} {Direct measurement of light waves},}\ }\href {https://doi.org/10.1126/science.1100866} {\bibfield  {journal} {\bibinfo  {journal} {Science}\ }\textbf {\bibinfo {volume} {305}},\ \bibinfo {pages} {1267--1269} (\bibinfo {year}
  {2004})}\BibitemShut {NoStop}%
\bibitem [{\citenamefont {Wirth}\ \emph {et~al.}(2011)\citenamefont {Wirth}, \citenamefont {Hassan}, \citenamefont {Grguraš}, \citenamefont {Gagnon}, \citenamefont {Moulet}, \citenamefont {Luu}, \citenamefont {Pabst}, \citenamefont {Santra}, \citenamefont {Alahmed}, \citenamefont {Azzeer},\ and\ \citenamefont {Krausz}}]{Wirth2011}%
  \BibitemOpen
  \bibfield  {author} {\bibinfo {author} {\bibfnamefont {A.}~\bibnamefont {Wirth}}, \bibinfo {author} {\bibfnamefont {M.~T.}\ \bibnamefont {Hassan}}, \bibinfo {author} {\bibfnamefont {I.}~\bibnamefont {Grguraš}}, \bibinfo {author} {\bibfnamefont {J.}~\bibnamefont {Gagnon}}, \bibinfo {author} {\bibfnamefont {A.}~\bibnamefont {Moulet}}, \bibinfo {author} {\bibfnamefont {T.~T.}\ \bibnamefont {Luu}}, \bibinfo {author} {\bibfnamefont {S.}~\bibnamefont {Pabst}}, \bibinfo {author} {\bibfnamefont {R.}~\bibnamefont {Santra}}, \bibinfo {author} {\bibfnamefont {Z.~A.}\ \bibnamefont {Alahmed}}, \bibinfo {author} {\bibfnamefont {A.~M.}\ \bibnamefont {Azzeer}},\ and\ \bibinfo {author} {\bibfnamefont {F.}~\bibnamefont {Krausz}},\ }\bibfield  {title} {\enquote {\bibinfo {title} {Synthesized light transients},}\ }\href {https://doi.org/10.1126/science.1210268} {\bibfield  {journal} {\bibinfo  {journal} {Science}\ }\textbf {\bibinfo {volume} {334}},\ \bibinfo {pages} {195--200} (\bibinfo {year} {2011})}\BibitemShut {NoStop}%
\bibitem [{\citenamefont {Hassan}\ \emph {et~al.}(2012)\citenamefont {Hassan}, \citenamefont {Wirth}, \citenamefont {Grguraš}, \citenamefont {Moulet}, \citenamefont {Luu}, \citenamefont {Gagnon}, \citenamefont {Pervak},\ and\ \citenamefont {Goulielmakis}}]{Hassan2012}%
  \BibitemOpen
  \bibfield  {author} {\bibinfo {author} {\bibfnamefont {M.~T.}\ \bibnamefont {Hassan}}, \bibinfo {author} {\bibfnamefont {A.}~\bibnamefont {Wirth}}, \bibinfo {author} {\bibfnamefont {I.}~\bibnamefont {Grguraš}}, \bibinfo {author} {\bibfnamefont {A.}~\bibnamefont {Moulet}}, \bibinfo {author} {\bibfnamefont {T.~T.}\ \bibnamefont {Luu}}, \bibinfo {author} {\bibfnamefont {J.}~\bibnamefont {Gagnon}}, \bibinfo {author} {\bibfnamefont {V.}~\bibnamefont {Pervak}},\ and\ \bibinfo {author} {\bibfnamefont {E.}~\bibnamefont {Goulielmakis}},\ }\bibfield  {title} {\enquote {\bibinfo {title} {Invited article: attosecond photonics: synthesis and control of light transients},}\ }\href {https://doi.org/10.1063/1.4742770} {\bibfield  {journal} {\bibinfo  {journal} {Review of Scientific Instruments}\ }\textbf {\bibinfo {volume} {83}},\ \bibinfo {pages} {111301} (\bibinfo {year} {2012})}\BibitemShut {NoStop}%
\bibitem [{\citenamefont {Hassan}\ \emph {et~al.}(2016)\citenamefont {Hassan}, \citenamefont {Luu}, \citenamefont {Moulet}, \citenamefont {Raskazovskaya}, \citenamefont {Zhokhov}, \citenamefont {Garg}, \citenamefont {Karpowicz}, \citenamefont {Zheltikov}, \citenamefont {Pervak}, \citenamefont {Krausz},\ and\ \citenamefont {Goulielmakis}}]{Hassan2016}%
  \BibitemOpen
  \bibfield  {author} {\bibinfo {author} {\bibfnamefont {M.~T.}\ \bibnamefont {Hassan}}, \bibinfo {author} {\bibfnamefont {T.~T.}\ \bibnamefont {Luu}}, \bibinfo {author} {\bibfnamefont {A.}~\bibnamefont {Moulet}}, \bibinfo {author} {\bibfnamefont {O.}~\bibnamefont {Raskazovskaya}}, \bibinfo {author} {\bibfnamefont {P.}~\bibnamefont {Zhokhov}}, \bibinfo {author} {\bibfnamefont {M.}~\bibnamefont {Garg}}, \bibinfo {author} {\bibfnamefont {N.}~\bibnamefont {Karpowicz}}, \bibinfo {author} {\bibfnamefont {A.~M.}\ \bibnamefont {Zheltikov}}, \bibinfo {author} {\bibfnamefont {V.}~\bibnamefont {Pervak}}, \bibinfo {author} {\bibfnamefont {F.}~\bibnamefont {Krausz}},\ and\ \bibinfo {author} {\bibfnamefont {E.}~\bibnamefont {Goulielmakis}},\ }\bibfield  {title} {\enquote {\bibinfo {title} {Optical attosecond pulses and tracking the nonlinear response of bound electrons},}\ }\href {https://doi.org/10.1038/nature16528} {\bibfield  {journal} {\bibinfo  {journal} {Nature}\ }\textbf {\bibinfo {volume} {530}},\ \bibinfo {pages}
  {66--70} (\bibinfo {year} {2016})}\BibitemShut {NoStop}%
\bibitem [{\citenamefont {Sweetser}, \citenamefont {Fittinghoff},\ and\ \citenamefont {Trebino}(1997)}]{Sweetser1997}%
  \BibitemOpen
  \bibfield  {author} {\bibinfo {author} {\bibfnamefont {J.~N.}\ \bibnamefont {Sweetser}}, \bibinfo {author} {\bibfnamefont {D.~N.}\ \bibnamefont {Fittinghoff}},\ and\ \bibinfo {author} {\bibfnamefont {R.}~\bibnamefont {Trebino}},\ }\bibfield  {title} {\enquote {\bibinfo {title} {Transient-grating frequency-resolved optical gating},}\ }\href {https://doi.org/10.1364/OL.22.000519} {\bibfield  {journal} {\bibinfo  {journal} {Optics Letters}\ }\textbf {\bibinfo {volume} {22}},\ \bibinfo {pages} {519--521} (\bibinfo {year} {1997})}\BibitemShut {NoStop}%
\bibitem [{\citenamefont {Nefedova}\ \emph {et~al.}(2018)\citenamefont {Nefedova}, \citenamefont {Ciappina}, \citenamefont {Finke}, \citenamefont {Albrecht}, \citenamefont {Kozlová},\ and\ \citenamefont {Nejdl}}]{Nefedova2018}%
  \BibitemOpen
  \bibfield  {author} {\bibinfo {author} {\bibfnamefont {V.~E.}\ \bibnamefont {Nefedova}}, \bibinfo {author} {\bibfnamefont {M.~F.}\ \bibnamefont {Ciappina}}, \bibinfo {author} {\bibfnamefont {O.}~\bibnamefont {Finke}}, \bibinfo {author} {\bibfnamefont {M.}~\bibnamefont {Albrecht}}, \bibinfo {author} {\bibfnamefont {M.}~\bibnamefont {Kozlová}},\ and\ \bibinfo {author} {\bibfnamefont {J.}~\bibnamefont {Nejdl}},\ }\bibfield  {title} {\enquote {\bibinfo {title} {Efficiency control of high-order harmonic generation in gases using driving pulse spectral features},}\ }\href {https://doi.org/10.1063/1.5050595} {\bibfield  {journal} {\bibinfo  {journal} {Applied Physics Letters}\ }\textbf {\bibinfo {volume} {113}},\ \bibinfo {pages} {191104} (\bibinfo {year} {2018})}\BibitemShut {NoStop}%
\bibitem [{\citenamefont {Chevreuil}\ \emph {et~al.}(2021)\citenamefont {Chevreuil}, \citenamefont {Brunner}, \citenamefont {Hrisafov}, \citenamefont {Pupeikis}, \citenamefont {Phillips}, \citenamefont {Keller},\ and\ \citenamefont {Gallmann}}]{Chevreuil2021}%
  \BibitemOpen
  \bibfield  {author} {\bibinfo {author} {\bibfnamefont {P.~A.}\ \bibnamefont {Chevreuil}}, \bibinfo {author} {\bibfnamefont {F.}~\bibnamefont {Brunner}}, \bibinfo {author} {\bibfnamefont {S.}~\bibnamefont {Hrisafov}}, \bibinfo {author} {\bibfnamefont {J.}~\bibnamefont {Pupeikis}}, \bibinfo {author} {\bibfnamefont {C.~R.}\ \bibnamefont {Phillips}}, \bibinfo {author} {\bibfnamefont {U.}~\bibnamefont {Keller}},\ and\ \bibinfo {author} {\bibfnamefont {L.}~\bibnamefont {Gallmann}},\ }\bibfield  {title} {\enquote {\bibinfo {title} {Water-window high harmonic generation with 0.8-{$\mu$}m and 2.2-{$\mu$}m {OPCPAs} at 100 k{Hz}},}\ }\href {https://doi.org/10.1364/OE.433649} {\bibfield  {journal} {\bibinfo  {journal} {Optics Express}\ }\textbf {\bibinfo {volume} {29}},\ \bibinfo {pages} {32996--33008} (\bibinfo {year} {2021})}\BibitemShut {NoStop}%
\bibitem [{\citenamefont {Kita}\ \emph {et~al.}(1983)\citenamefont {Kita}, \citenamefont {Harada}, \citenamefont {Nakano},\ and\ \citenamefont {Kuroda}}]{Kita1983}%
  \BibitemOpen
  \bibfield  {author} {\bibinfo {author} {\bibfnamefont {T.}~\bibnamefont {Kita}}, \bibinfo {author} {\bibfnamefont {T.}~\bibnamefont {Harada}}, \bibinfo {author} {\bibfnamefont {N.}~\bibnamefont {Nakano}},\ and\ \bibinfo {author} {\bibfnamefont {H.}~\bibnamefont {Kuroda}},\ }\bibfield  {title} {\enquote {\bibinfo {title} {Mechanically ruled aberration-corrected concave gratings for a flat-field grazing-incidence spectrograph},}\ }\href {https://doi.org/10.1364/AO.22.000512} {\bibfield  {journal} {\bibinfo  {journal} {Applied Optics}\ }\textbf {\bibinfo {volume} {22}},\ \bibinfo {pages} {512--513} (\bibinfo {year} {1983})}\BibitemShut {NoStop}%
\bibitem [{\citenamefont {Harada}\ \emph {et~al.}(1999)\citenamefont {Harada}, \citenamefont {Takahashi}, \citenamefont {Sakuma},\ and\ \citenamefont {Osyczka}}]{Harada1999}%
  \BibitemOpen
  \bibfield  {author} {\bibinfo {author} {\bibfnamefont {T.}~\bibnamefont {Harada}}, \bibinfo {author} {\bibfnamefont {K.}~\bibnamefont {Takahashi}}, \bibinfo {author} {\bibfnamefont {H.}~\bibnamefont {Sakuma}},\ and\ \bibinfo {author} {\bibfnamefont {A.}~\bibnamefont {Osyczka}},\ }\bibfield  {title} {\enquote {\bibinfo {title} {Optimum design of a grazing-incidence flat-field spectrograph with a spherical varied-line-space grating},}\ }\href {https://doi.org/10.1364/AO.38.002743} {\bibfield  {journal} {\bibinfo  {journal} {Applied Optics}\ }\textbf {\bibinfo {volume} {38}},\ \bibinfo {pages} {2743--2748} (\bibinfo {year} {1999})}\BibitemShut {NoStop}%
\bibitem [{\citenamefont {Rosenfeld}\ \emph {et~al.}(1999)\citenamefont {Rosenfeld}, \citenamefont {Lorenz}, \citenamefont {Stoian},\ and\ \citenamefont {Ashkenasi}}]{Rosenfeld1999}%
  \BibitemOpen
  \bibfield  {author} {\bibinfo {author} {\bibfnamefont {A.}~\bibnamefont {Rosenfeld}}, \bibinfo {author} {\bibfnamefont {M.}~\bibnamefont {Lorenz}}, \bibinfo {author} {\bibfnamefont {R.}~\bibnamefont {Stoian}},\ and\ \bibinfo {author} {\bibfnamefont {D.}~\bibnamefont {Ashkenasi}},\ }\bibfield  {title} {\enquote {\bibinfo {title} {Ultrashort-laser-pulse damage threshold of transparent materials and the role of incubation},}\ }\href {https://doi.org/10.1007/s003390051419} {\bibfield  {journal} {\bibinfo  {journal} {Applied Physics A}\ }\textbf {\bibinfo {volume} {69}},\ \bibinfo {pages} {S373--S376} (\bibinfo {year} {1999})}\BibitemShut {NoStop}%
\bibitem [{\citenamefont {Lenzner}\ \emph {et~al.}(1999)\citenamefont {Lenzner}, \citenamefont {Kruger}, \citenamefont {Kautek},\ and\ \citenamefont {Krausz}}]{Lenzner1999}%
  \BibitemOpen
  \bibfield  {author} {\bibinfo {author} {\bibfnamefont {M.}~\bibnamefont {Lenzner}}, \bibinfo {author} {\bibfnamefont {J.}~\bibnamefont {Kruger}}, \bibinfo {author} {\bibfnamefont {W.}~\bibnamefont {Kautek}},\ and\ \bibinfo {author} {\bibfnamefont {F.}~\bibnamefont {Krausz}},\ }\bibfield  {title} {\enquote {\bibinfo {title} {Incubation of laser ablation in fused silica with 5-fs pulses},}\ }\href {https://doi.org/10.1007/s003390051034} {\bibfield  {journal} {\bibinfo  {journal} {Applied Physics A}\ }\textbf {\bibinfo {volume} {69}},\ \bibinfo {pages} {465--466} (\bibinfo {year} {1999})}\BibitemShut {NoStop}%
\bibitem [{\citenamefont {Smalakys}\ \emph {et~al.}(2019)\citenamefont {Smalakys}, \citenamefont {Momgaudis}, \citenamefont {Grigutis}, \citenamefont {Kičas},\ and\ \citenamefont {Melninkaitis}}]{Smalakys2019}%
  \BibitemOpen
  \bibfield  {author} {\bibinfo {author} {\bibfnamefont {L.}~\bibnamefont {Smalakys}}, \bibinfo {author} {\bibfnamefont {B.}~\bibnamefont {Momgaudis}}, \bibinfo {author} {\bibfnamefont {R.}~\bibnamefont {Grigutis}}, \bibinfo {author} {\bibfnamefont {S.}~\bibnamefont {Kičas}},\ and\ \bibinfo {author} {\bibfnamefont {A.}~\bibnamefont {Melninkaitis}},\ }\bibfield  {title} {\enquote {\bibinfo {title} {Contrasted fatigue behavior of laser-induced damage mechanisms in single layer {ZrO$_2$} optical coating},}\ }\href {https://doi.org/10.1364/OE.27.026088} {\bibfield  {journal} {\bibinfo  {journal} {Optics Express}\ }\textbf {\bibinfo {volume} {27}},\ \bibinfo {pages} {26088--26101} (\bibinfo {year} {2019})}\BibitemShut {NoStop}%
\bibitem [{\citenamefont {M\"oller}\ \emph {et~al.}(2007)\citenamefont {M\"oller}, \citenamefont {Andresen}, \citenamefont {Merschjann}, \citenamefont {Zimmermann}, \citenamefont {Prinz},\ and\ \citenamefont {Imlau}}]{Moller2007}%
  \BibitemOpen
  \bibfield  {author} {\bibinfo {author} {\bibfnamefont {S.}~\bibnamefont {M\"oller}}, \bibinfo {author} {\bibfnamefont {A.}~\bibnamefont {Andresen}}, \bibinfo {author} {\bibfnamefont {C.}~\bibnamefont {Merschjann}}, \bibinfo {author} {\bibfnamefont {B.}~\bibnamefont {Zimmermann}}, \bibinfo {author} {\bibfnamefont {M.}~\bibnamefont {Prinz}},\ and\ \bibinfo {author} {\bibfnamefont {M.}~\bibnamefont {Imlau}},\ }\bibfield  {title} {\enquote {\bibinfo {title} {Insight to {UV}-induced formation of laser damage on {LiB$_3$O$_5$} optical surfaces during long-term sum-frequency generation},}\ }\href {https://doi.org/10.1364/OE.15.007351} {\bibfield  {journal} {\bibinfo  {journal} {Optics Express}\ }\textbf {\bibinfo {volume} {15}},\ \bibinfo {pages} {7351--7356} (\bibinfo {year} {2007})}\BibitemShut {NoStop}%
\bibitem [{\citenamefont {Hong}\ \emph {et~al.}(2013)\citenamefont {Hong}, \citenamefont {Liu}, \citenamefont {Huang},\ and\ \citenamefont {Gong}}]{Hong2013}%
  \BibitemOpen
  \bibfield  {author} {\bibinfo {author} {\bibfnamefont {H.}~\bibnamefont {Hong}}, \bibinfo {author} {\bibfnamefont {Q.}~\bibnamefont {Liu}}, \bibinfo {author} {\bibfnamefont {L.}~\bibnamefont {Huang}},\ and\ \bibinfo {author} {\bibfnamefont {M.}~\bibnamefont {Gong}},\ }\bibfield  {title} {\enquote {\bibinfo {title} {Improvement and formation of {UV}-induced damage on {LBO} crystal surface during long-term high-power third-harmonic generation},}\ }\href {https://doi.org/10.1364/OE.21.007285} {\bibfield  {journal} {\bibinfo  {journal} {Optics Express}\ }\textbf {\bibinfo {volume} {21}},\ \bibinfo {pages} {7285--7293} (\bibinfo {year} {2013})}\BibitemShut {NoStop}%
\bibitem [{\citenamefont {Vampa}\ \emph {et~al.}(2017)\citenamefont {Vampa}, \citenamefont {Ghamsari}, \citenamefont {Siadat~Mousavi}, \citenamefont {Hammond}, \citenamefont {Olivieri}, \citenamefont {Lisicka-Skrek}, \citenamefont {Naumov}, \citenamefont {Villeneuve}, \citenamefont {Staudte}, \citenamefont {Berini},\ and\ \citenamefont {Corkum}}]{Vampa2017}%
  \BibitemOpen
  \bibfield  {author} {\bibinfo {author} {\bibfnamefont {G.}~\bibnamefont {Vampa}}, \bibinfo {author} {\bibfnamefont {B.~G.}\ \bibnamefont {Ghamsari}}, \bibinfo {author} {\bibfnamefont {S.}~\bibnamefont {Siadat~Mousavi}}, \bibinfo {author} {\bibfnamefont {T.~J.}\ \bibnamefont {Hammond}}, \bibinfo {author} {\bibfnamefont {A.}~\bibnamefont {Olivieri}}, \bibinfo {author} {\bibfnamefont {E.}~\bibnamefont {Lisicka-Skrek}}, \bibinfo {author} {\bibfnamefont {A.~Y.}\ \bibnamefont {Naumov}}, \bibinfo {author} {\bibfnamefont {D.~M.}\ \bibnamefont {Villeneuve}}, \bibinfo {author} {\bibfnamefont {A.}~\bibnamefont {Staudte}}, \bibinfo {author} {\bibfnamefont {P.}~\bibnamefont {Berini}},\ and\ \bibinfo {author} {\bibfnamefont {P.~B.}\ \bibnamefont {Corkum}},\ }\bibfield  {title} {\enquote {\bibinfo {title} {Plasmon-enhanced high-harmonic generation from silicon},}\ }\href {https://doi.org/10.1038/nphys4087} {\bibfield  {journal} {\bibinfo  {journal} {Nature Physics}\ }\textbf {\bibinfo {volume} {13}},\ \bibinfo {pages}
  {659--662} (\bibinfo {year} {2017})}\BibitemShut {NoStop}%
\bibitem [{\citenamefont {Huttner}, \citenamefont {Kira},\ and\ \citenamefont {Koch}(2017)}]{Huttner2017}%
  \BibitemOpen
  \bibfield  {author} {\bibinfo {author} {\bibfnamefont {U.}~\bibnamefont {Huttner}}, \bibinfo {author} {\bibfnamefont {M.}~\bibnamefont {Kira}},\ and\ \bibinfo {author} {\bibfnamefont {S.~W.}\ \bibnamefont {Koch}},\ }\bibfield  {title} {\enquote {\bibinfo {title} {Ultrahigh {O}ff-{R}esonant {F}ield {E}ffects in {S}emiconductors},}\ }\href {https://doi.org/10.1002/lpor.201700049} {\bibfield  {journal} {\bibinfo  {journal} {Laser \& Photonics Reviews}\ }\textbf {\bibinfo {volume} {11}},\ \bibinfo {pages} {1700049} (\bibinfo {year} {2017})}\BibitemShut {NoStop}%
\bibitem [{\citenamefont {Liu}\ \emph {et~al.}(2018)\citenamefont {Liu}, \citenamefont {Guo}, \citenamefont {Vampa}, \citenamefont {Zhang}, \citenamefont {Sarmiento}, \citenamefont {Xiao}, \citenamefont {Bucksbaum}, \citenamefont {Vučkovič}, \citenamefont {Fan},\ and\ \citenamefont {Reis}}]{Liu2018}%
  \BibitemOpen
  \bibfield  {author} {\bibinfo {author} {\bibfnamefont {H.}~\bibnamefont {Liu}}, \bibinfo {author} {\bibfnamefont {C.}~\bibnamefont {Guo}}, \bibinfo {author} {\bibfnamefont {G.}~\bibnamefont {Vampa}}, \bibinfo {author} {\bibfnamefont {J.~L.}\ \bibnamefont {Zhang}}, \bibinfo {author} {\bibfnamefont {T.}~\bibnamefont {Sarmiento}}, \bibinfo {author} {\bibfnamefont {M.}~\bibnamefont {Xiao}}, \bibinfo {author} {\bibfnamefont {P.~H.}\ \bibnamefont {Bucksbaum}}, \bibinfo {author} {\bibfnamefont {J.}~\bibnamefont {Vučkovič}}, \bibinfo {author} {\bibfnamefont {S.}~\bibnamefont {Fan}},\ and\ \bibinfo {author} {\bibfnamefont {D.~A.}\ \bibnamefont {Reis}},\ }\bibfield  {title} {\enquote {\bibinfo {title} {Enhanced high-harmonic generation from an all-dielectric metasurface},}\ }\href {https://doi.org/10.1038/s41567-018-0233-6} {\bibfield  {journal} {\bibinfo  {journal} {Nature Physics}\ }\textbf {\bibinfo {volume} {14}},\ \bibinfo {pages} {1006--1010} (\bibinfo {year} {2018})}\BibitemShut {NoStop}%
\bibitem [{\citenamefont {Goulielmakis}\ \emph {et~al.}(2008)\citenamefont {Goulielmakis}, \citenamefont {Schultze}, \citenamefont {Hofstetter}, \citenamefont {Yakovlev}, \citenamefont {Gagnon}, \citenamefont {Uiberacker}, \citenamefont {Aquila}, \citenamefont {Gullikson}, \citenamefont {Attwood}, \citenamefont {Kienberger},\ and\ \citenamefont {Krausz}}]{Goulielmakis2008}%
  \BibitemOpen
  \bibfield  {author} {\bibinfo {author} {\bibfnamefont {E.}~\bibnamefont {Goulielmakis}}, \bibinfo {author} {\bibfnamefont {M.}~\bibnamefont {Schultze}}, \bibinfo {author} {\bibfnamefont {M.}~\bibnamefont {Hofstetter}}, \bibinfo {author} {\bibfnamefont {V.~S.}\ \bibnamefont {Yakovlev}}, \bibinfo {author} {\bibfnamefont {J.}~\bibnamefont {Gagnon}}, \bibinfo {author} {\bibfnamefont {M.}~\bibnamefont {Uiberacker}}, \bibinfo {author} {\bibfnamefont {A.~L.}\ \bibnamefont {Aquila}}, \bibinfo {author} {\bibfnamefont {E.~M.}\ \bibnamefont {Gullikson}}, \bibinfo {author} {\bibfnamefont {D.~T.}\ \bibnamefont {Attwood}}, \bibinfo {author} {\bibfnamefont {R.}~\bibnamefont {Kienberger}},\ and\ \bibinfo {author} {\bibfnamefont {F.}~\bibnamefont {Krausz}},\ }\bibfield  {title} {\enquote {\bibinfo {title} {Single-cycle nonlinear optics},}\ }\href {https://doi.org/10.1126/science.1157846} {\bibfield  {journal} {\bibinfo  {journal} {Science}\ }\textbf {\bibinfo {volume} {320}},\ \bibinfo {pages} {1614--1617} (\bibinfo {year}
  {2008})}\BibitemShut {NoStop}%
\bibitem [{\citenamefont {Schultze}\ \emph {et~al.}(2011)\citenamefont {Schultze}, \citenamefont {Wirth}, \citenamefont {Grguras}, \citenamefont {Uiberacker}, \citenamefont {Uphues}, \citenamefont {Verhoef}, \citenamefont {Gagnon}, \citenamefont {Hofstetter}, \citenamefont {Kleineberg}, \citenamefont {Goulielmakis},\ and\ \citenamefont {Krausz}}]{Schultze2011}%
  \BibitemOpen
  \bibfield  {author} {\bibinfo {author} {\bibfnamefont {M.}~\bibnamefont {Schultze}}, \bibinfo {author} {\bibfnamefont {A.}~\bibnamefont {Wirth}}, \bibinfo {author} {\bibfnamefont {I.}~\bibnamefont {Grguras}}, \bibinfo {author} {\bibfnamefont {M.}~\bibnamefont {Uiberacker}}, \bibinfo {author} {\bibfnamefont {T.}~\bibnamefont {Uphues}}, \bibinfo {author} {\bibfnamefont {A.~J.}\ \bibnamefont {Verhoef}}, \bibinfo {author} {\bibfnamefont {J.}~\bibnamefont {Gagnon}}, \bibinfo {author} {\bibfnamefont {M.}~\bibnamefont {Hofstetter}}, \bibinfo {author} {\bibfnamefont {U.}~\bibnamefont {Kleineberg}}, \bibinfo {author} {\bibfnamefont {E.}~\bibnamefont {Goulielmakis}},\ and\ \bibinfo {author} {\bibfnamefont {F.}~\bibnamefont {Krausz}},\ }\bibfield  {title} {\enquote {\bibinfo {title} {State-of-the-art attosecond metrology},}\ }\href {https://doi.org/10.1016/j.elspec.2010.12.015} {\bibfield  {journal} {\bibinfo  {journal} {Journal of Electron Spectroscopy and Related Phenomena}\ }\textbf {\bibinfo {volume} {184}},\
  \bibinfo {pages} {68--77} (\bibinfo {year} {2011})}\BibitemShut {NoStop}%
\bibitem [{\citenamefont {Hussain}\ \emph {et~al.}(2020)\citenamefont {Hussain}, \citenamefont {Pires}, \citenamefont {Boutu}, \citenamefont {Franz}, \citenamefont {Nicolas}, \citenamefont {Imran}, \citenamefont {Merdji}, \citenamefont {Fajardo},\ and\ \citenamefont {Williams}}]{Hussain2020}%
  \BibitemOpen
  \bibfield  {author} {\bibinfo {author} {\bibfnamefont {M.}~\bibnamefont {Hussain}}, \bibinfo {author} {\bibfnamefont {H.}~\bibnamefont {Pires}}, \bibinfo {author} {\bibfnamefont {W.}~\bibnamefont {Boutu}}, \bibinfo {author} {\bibfnamefont {D.}~\bibnamefont {Franz}}, \bibinfo {author} {\bibfnamefont {R.}~\bibnamefont {Nicolas}}, \bibinfo {author} {\bibfnamefont {T.}~\bibnamefont {Imran}}, \bibinfo {author} {\bibfnamefont {H.}~\bibnamefont {Merdji}}, \bibinfo {author} {\bibfnamefont {M.}~\bibnamefont {Fajardo}},\ and\ \bibinfo {author} {\bibfnamefont {G.~O.}\ \bibnamefont {Williams}},\ }\bibfield  {title} {\enquote {\bibinfo {title} {Controlling the non-linear optical properties of {MgO} by tailoring the electronic structure},}\ }\href {https://doi.org/10.1007/s00340-020-6360-3} {\bibfield  {journal} {\bibinfo  {journal} {Applied Physics B}\ }\textbf {\bibinfo {volume} {126}},\ \bibinfo {pages} {46} (\bibinfo {year} {2020})}\BibitemShut {NoStop}%
\bibitem [{\citenamefont {Heinrich}\ \emph {et~al.}(2021)\citenamefont {Heinrich}, \citenamefont {Taucer}, \citenamefont {Kfir}, \citenamefont {Corkum}, \citenamefont {Staudte}, \citenamefont {Ropers},\ and\ \citenamefont {Sivis}}]{Heinrich2021}%
  \BibitemOpen
  \bibfield  {author} {\bibinfo {author} {\bibfnamefont {T.}~\bibnamefont {Heinrich}}, \bibinfo {author} {\bibfnamefont {M.}~\bibnamefont {Taucer}}, \bibinfo {author} {\bibfnamefont {O.}~\bibnamefont {Kfir}}, \bibinfo {author} {\bibfnamefont {P.~B.}\ \bibnamefont {Corkum}}, \bibinfo {author} {\bibfnamefont {A.}~\bibnamefont {Staudte}}, \bibinfo {author} {\bibfnamefont {C.}~\bibnamefont {Ropers}},\ and\ \bibinfo {author} {\bibfnamefont {M.}~\bibnamefont {Sivis}},\ }\bibfield  {title} {\enquote {\bibinfo {title} {Chiral high-harmonic generation and spectroscopy on solid surfaces using polarization-tailored strong fields},}\ }\href {https://doi.org/10.1038/s41467-021-23999-9} {\bibfield  {journal} {\bibinfo  {journal} {Nature Communications}\ }\textbf {\bibinfo {volume} {12}},\ \bibinfo {pages} {3723} (\bibinfo {year} {2021})}\BibitemShut {NoStop}%
\bibitem [{\citenamefont {Korobenko}\ \emph {et~al.}(2021)\citenamefont {Korobenko}, \citenamefont {Saha}, \citenamefont {Godfrey}, \citenamefont {Gertsvolf}, \citenamefont {Naumov}, \citenamefont {Villeneuve}, \citenamefont {Boltasseva}, \citenamefont {Shalaev},\ and\ \citenamefont {Corkum}}]{Korobenko2021}%
  \BibitemOpen
  \bibfield  {author} {\bibinfo {author} {\bibfnamefont {A.}~\bibnamefont {Korobenko}}, \bibinfo {author} {\bibfnamefont {S.}~\bibnamefont {Saha}}, \bibinfo {author} {\bibfnamefont {A.~T.~K.}\ \bibnamefont {Godfrey}}, \bibinfo {author} {\bibfnamefont {M.}~\bibnamefont {Gertsvolf}}, \bibinfo {author} {\bibfnamefont {A.~Y.}\ \bibnamefont {Naumov}}, \bibinfo {author} {\bibfnamefont {D.~M.}\ \bibnamefont {Villeneuve}}, \bibinfo {author} {\bibfnamefont {A.}~\bibnamefont {Boltasseva}}, \bibinfo {author} {\bibfnamefont {V.~M.}\ \bibnamefont {Shalaev}},\ and\ \bibinfo {author} {\bibfnamefont {P.~B.}\ \bibnamefont {Corkum}},\ }\bibfield  {title} {\enquote {\bibinfo {title} {High-harmonic generation in metallic titanium nitride},}\ }\href {https://doi.org/10.1038/s41467-021-24286-3} {\bibfield  {journal} {\bibinfo  {journal} {Nature Communications}\ }\textbf {\bibinfo {volume} {12}},\ \bibinfo {pages} {4981} (\bibinfo {year} {2021})}\BibitemShut {NoStop}%
\end{thebibliography}%
\end{document}